\documentclass[journal,10pt, a4paper,twocolumns,final]{IEEEtran}

\newlength{\flexwidth}
\usepackage{amsmath}
\usepackage{amssymb}
\usepackage{fixmath}
\usepackage{xcolor}
\usepackage{amsthm}
\usepackage{siunitx}
\usepackage{cuted}
\usepackage{multirow}
\usepackage{booktabs}
\usepackage{subcaption}
\usepackage[inline]{enumitem}
\usepackage{optidef}
\usepackage{graphicx}
\usepackage[norelsize,linesnumbered]{algorithm2e}
\makeatletter
\newcommand{\removelatexerror} {\let\@latex@error\@gobble}
\makeatother

\usepackage{tcolorbox}
\tcbuselibrary{many}

\newcommand{\superscript}[1]{^{\mathrm{#1}}}
\newcommand{\subscript}[1]{_{\mathrm{#1}}}

\newcommand{\diff}{\text{d}}

\newcommand{\revise}[1]{{\color{\highlightcolor}#1}}
\newcommand{\highlightcolor}{red}
\renewcommand{\highlightcolor}{black}
\newtheorem{theorem}{Theorem}
\newtheorem{lemma}{Lemma}
\newtheorem{corollary}{Corollary}

\newtcbtheorem[]{alg}{Algorithm}{fonttitle=\bfseries}{alg}

\begin{document}

\title{CLARQ: A Dynamic ARQ Solution for Ultra-high Closed-loop Reliability}


 \author{Bin~Han,
		Yao~Zhu,
		Muxia~Sun,
 		Vincenzo~Sciancalepore,
 		Yulin~Hu,
	 	and Hans~D.~Schotten
	 	\thanks{B. Han and H. D. Schotten are with the Division of Wireless Communications and Radio Positioning (WiCoN), University of Kaiserslautern, 67663 Kaiserslautern, Germany, Emails: \{binhan, schotten\}@eit.uni-kl.de. Y. Zhu and Y. Hu are with School of Electronic Information, Wuhan University,  430072 Wuhan, China,  and   with ISEK Research~Area, RWTH Aachen University, D-52074 Aachen. Germany,  Emails: \{yao.zhu, yulin.hu\}@isek.rwth-aachen.de. M. Sun is with the Department of Industrial Engineering, Tsinghua University, Beijing 100084, China, Email: muxiasun@mail.tsinghua.edu.cn. V. Sciancalepore is with NEC Laboratories Europe, 69115 Heidelberg, Germany, Email: vincenzo.sciancalepore@neclab.eu.}
 }


\maketitle

\begin{abstract}
%
	Emerging wireless control applications demand for extremely high closed-loop reliability under strict latency constraints, which the conventional Automatic Repeat reQuest (ARQ) solutions with static schedules fail to provide.	To overcome this issue and enable data-link layer error control for ultra reliable low-latency communication (URLLC) services, we propose a novel protocol: the Closed-Loop ARQ (CLARQ), which forces to accomplish an information exchange round within a fixed loop-back latency, and dynamically re-allocates the remaining resource between uplink and downlink slots upon the result of last uplink transmission. The proposed method guarantees to meet the latency requirement, \revise{while delivering} high communication reliability and power efficiency. It can be efficiently offline optimized by means of dynamic programming techniques, and is capable of real-time deployment with a low-cost implementation based on look-up tables. Numerical evaluations have verified that CLARQ outperforms baselines with significantly improved closed-loop reliability and reduced energy consumption. Especially, over a Rayleigh channel with \SI{0}{\dB} mean SNR, it is able to provide a closed-loop error rate below $10^{-7}$ within \SI{10}{\milli\second} loop-back latency, which makes our proposal competitive for practical URLLC applications in future 5G-and-beyond networks.
\end{abstract}


\IEEEpeerreviewmaketitle

\section{Introduction}

{The upcoming 5G-and-beyond and 6G network design is expected to fully support ultra-reliable low-latency communication (URLLC) services that will offer unprecedented market opportunities thereby attracting new business players. This would result in an extreme link reliability that represents an essential added-value for emerging wireless communication systems.} 
Advanced use cases such as railway communications, factory \& process automation, and autonomous driving~\cite{3GPP_TS_22.104_g20,3GPP_TS_22.886_g20,3GPP_TS_22.289_g10} will demand ultra-reliable communications to construct an innovative ecosystem that brings high reactiveness and strong reliability to existing network deployments. However, such requirements involve a number of technical challenges {to be carefully evaluated while pioneering novel} technologies~\cite{Bennis2018ultrareliable}.

While end-to-end (E2E) latency has been exhaustively addressed in the last few years with innovative technical actions~\cite{Jiang2019low,Zanzi2020laco} that usually rely on the tactile internet use-case requirements~\cite{Simsek20165g}, they lack of practical solutions that keep the service reliability at reasonable levels. 
{Applying Automatic Repeat reQuest (ARQ) or its hybrid version (HARQ) on the data link layer has been the most promising approach during the last decade allowing resending data packets that are not successfully delivered to the receiver.
Nonetheless, ARQ or HARQ methods might appear inadequate when low-latency scenarios are in place due to the following reasons}: $i$) while exploiting extra-time resources to repeat messages upon transmission failures, it significantly increases the E2E latency that in turn leads to a raised probability of violating latency requirements, as demonstrated by~\cite{Bennis2018ultrareliable} and $ii$) adding static ARQ/HARQ scheduling solutions that create retransmission slots within a frame of limited blocklength might bring no tangible gains but only loss to the link reliability, as also proven by the authors of~\cite{Makki2014finite}.
{Indeed, available solutions commonly rely on either radio resources scheduling among different users pursuing fairness maximization, or spatial/frequency diversity over different paths/channels to fulfill the reliability requirements:} both approaches require a significant complexity of the network, and may result in a spectral and power efficiency reduction.
Furthermore, many applications of ultra-reliable communications, such as automated control, are working in closed-loops, where the utility of a downlink (DL) transmission relies on a successful uplink (UL) transmission. However, most conventional solutions are designed to improve the open-loop link reliability, and fail to leverage this duplex asymmetry for better radio resource efficiency.

{In order to overcome above-mentioned issues, in this study we propose a novel ARQ-based protocol} that works in the finite blocklength (FBL) regime, where the resources pre-dedicated to the DL slots can be dynamically re-allocated and exploited for retransmission in UL upon failures. {Our proposal can automatically enable ARQ/HARQ techniques} in scenarios where the E2E latency is strictly limited, and therewith significantly increase the closed-loop reliability of communication that may appear especially critical when ultra-reliable use cases are in place.
{Various approaches have been already proposed in literature to fulfill the reliability requirements of URLLC-based use cases: for e.g. applying advanced resource allocation methods such that the radio resources can be more-efficiently shared $i$) among devices of different classes~\cite{Abedin2019resource}, and $ii$) among different URLLC data packets~\cite{Anand2018resource}.} In addition, it has been demonstrated that adaptive sub-carrier selection can also improve the link reliability in OFDM systems by raising the SNR and reducing adjacent-channel interference~\cite{Hamamreh2017ofdm}. \revise{Furthermore, demonstrated since long as effective to achieve a flexible trade-off between power and latency~\cite{Berry2000power}, adaptive power control over fading channels can be considered as a promising solution for URLLC}.

Differing from such physical layer approaches, {in this paper we propose a novel protocol, namely CLARQ that $i$) works in the finite blocklength (FBL) regime, where the resources pre-dedicated to the DL slots can be dynamically re-allocated and exploited for retransmission in UL upon failures, $ii$) can automatically enable ARQ/HARQ techniques in URLLC scenarios where the E2E latency is strictly limited, $iii$) can significantly increase the closed-loop reliability of communication that may appear especially critical when ultra-reliable use cases are in place, and $iv$) exploits the time diversity in an opportunistic fashion, and therefore is capable to apply to single-hop networks without making use of spatial-diversity-based methods, (e.g.~\cite{Mountaster2018reliable,Tseng2019selective,Luo2018frudp}).}

The residential contents of this paper are organized as follows: we {begin with a revisit to the well-studied} classic FBL problem of cross-user blocklength allocation in Section~\ref{sec:tdma_case}, to provide insights on the background knowledge of FBL information theory. Then, in Section~\ref{sec:static_closedloop} we setup the UL/DL blocklength allocation problem in {closed-loop communication} systems, showing the optimum and highlighting the limits of ARQ/HARQ mechanisms with static schedules. Section~\ref{sec:clarq} presents our main contributions, which consist of: $i$) a new protocol design to enable dynamic ARQ within limited blocklength, $ii$) an optimal policy analysis of a dynamic retransmission showing pros and cons and, $iii$) a dynamic programming algorithm to numerically derive the optimal policy. Section~\ref{sec:eval} presents an exhaustive simulation campaign to prove the validness of both the protocol and the optimizer in comparison with conventional benchmarks.{ Regarding practical implementation and deployment in realistic radio environments, in  Section~\ref{sec:discussions} we further extend our discussion to several aspects of technical details. To the end,} in Section~\ref{sec:related} we refer to related work on the topic, before Section~\ref{sec:conclusion} closes the paper with our conclusion and outlooks to future works.

\section{Preliminary Analysis: Cross-user Blocklength Allocation in TDMA}\label{sec:tdma_case}

Existing studies in the field of FBL transmission commonly focus on the blocklength allocation problem in TDMA systems, of which a typical case can be summarized as follows. Given $M$ devices $m\in\mathcal{M}\overset{\Delta}{=}\{1,2\dots M\}$ that share a time frame of $T$ to transmit their messages to the server, where all messages have the same bit length $d$; given an upper bound $\varepsilon\subscript{max}$ for the message error rate for every device, it searches the optimal allocation of time (blocklength) that maximizes the expected sum of successfully transmitted messages:
\begin{maxi!}[2]
	{\mathbf{n}\in\mathbb{N}^M}{\sum\limits_{m\in\mathcal{M}}^{}\left(1-\varepsilon_m\right)}{\label{prob:classic_fbl}}{}
	\addConstraint{\sum\limits_{m\in\mathcal{M}}n_mT\subscript{S}\leqslant T}
	\addConstraint{\varepsilon_m\le\varepsilon\subscript{max},\quad\forall m\in\mathcal{M}.}
\end{maxi!}
Here, $T\subscript{S}$ is the symbol length and $\mathbf{n}=[n_1,n_2\dots n_M]$ describes the blocklength allocation among devices. In the FBL regime, {according to \cite{Polyanskiy2010channel}}, the message error rate of device $m$ is
\begin{equation}
\varepsilon_m\approx Q\left(\sqrt{\frac{n_m}{V_m}}(\mathcal{C}_m-r_m)\ln{2}\right),
\label{eq:err_prob}
\end{equation}
where $r_m=d/n_m$ is the block coding rate of device $m$, $\mathcal{C}_m=\log_2(1+\gamma_m)$ the Shannon capacity of device $m$, $\gamma_m$ is the SNR at device $m$, and $V_m$ is the channel dispersion for device $m$, which equals $1-1/(1+\gamma_m)^2$ for complex AWGN channels. This classic FBL problem \eqref{prob:classic_fbl} is usually studied in its relaxed form where $\mathbf{n}\in{\mathbb{R}^+}^M$, which is proven convex. \revise{Without any retransmission scheme, the problem has an unique optimum $\mathbf{n}\subscript{opt}$, i.e., $n\subscript{opt,m}=\varepsilon^{-1}(\varepsilon\subscript{opt})$, where $\varepsilon^{(-1)}$ is the inverse function of (2) 
and $\sum\limits_{m\in\mathcal{M}}n\subscript{opt,m}=T/T\subscript{S}$ with $\varepsilon_m=\varepsilon\subscript{opt}$ for all $m\in\mathcal{M}$}{\footnotemark}. The integer solution to \eqref{prob:classic_fbl} can be approximated by rounding this $\mathbf{n}\subscript{opt}$~\cite{Zhu2019reliability}.
\footnotetext{{Remark that \eqref{eq:err_prob} applies only for AWGN channel assuming perfect CSI is available. Nevertheless, even in \revise{a} lack of perfect CSI, we have derived in another recent work~\cite{Han2020fairness} that the convexity of  $\varepsilon_m$ w.r.t. $n_m$ still holds as long as the statistical distribution of CSI is known. Besides, the expression of channel dispersion $V_m$ can be extended to the cases of Gaussian-mixture and generic non-Gaussian channels w.r.t the analyses in \cite{Polyanskiy2009dispersion} and \cite{Scarlett2017dispersion}, respectively. Such modifications in the form of $V_m$, however, do not deny \revise{any of} our analyses or proposals in this manuscript.}}
%

Futhermore, noticing in \eqref{eq:err_prob} that the message error probability $\varepsilon_m$ is monotonically decreasing w.r.t. $n_m$, while retransmission protocols such as Automatic Repeat reQuest (ARQ) can provide a gain of link reliability by dividing $n_m$ into several sub-slots, interests have been raised to study the retransmission problem in the FBL regime. In this problem, the time slot $n_m$ allocated to every device $m$ is further uniformly divided into $N$ sub-slots, and the device $m$ attempts to transmit its message to the server within a sub-slot. Upon message error, up to $N-1$ retransmissions are allowed for every device through an ARQ mechanism. For simplification, the feedback of Acknowledgment / Non-Acknowledgment (ACK/NACK) message is usually considered reliable and the feedback cost is neglected. Thus, the blocklength of every uplink transmission attempt by device $m$ is the sub-slot length, i.e. $n_m/N$.

Under these assumptions, the authors of~\cite{Zhu2019energy} have show that the total energy consumption with retransmission-enabled system in the edge computing network is more energy-efficient comparing to that with the one-shot scheme under the finite blocklength regime. The maximal number of transmission attempts $N$ and the frame structure can be optimized to minimize the overall energy consumption. However, in perspective of the message error rate minimization, it has been proven by ~\cite{Makki2014finite} that the minimal achievable error probability with Hybrid ARQ (HARQ) is equal to that with one-shot transmission. More specifically, for an arbitrary device allocated with a certain blocklength $N$, denote  by $\varepsilon_{(i)}$ the error probability up to $i^{th}$ re-/transmission attempts and $\varepsilon_{i}$ the error probability of the $i^{th}$ re-/transmission attempt, the overall message transmission error probability with HARQ is given by:
\begin{equation}
\begin{split}
\varepsilon_{(i)}&=\varepsilon_{(i-1)}+\varepsilon_{i}-\varepsilon_{(i-1)}\varepsilon_i\approx Q\left(\frac{\mathcal{C}-\frac{d}{in}}{\sqrt{\frac{V}{in}}}\right)
\end{split}
\label{eq:error_HARQ}
\end{equation}
where $n$ is the sub-slot blocklength. With a maximal transmission attempts $I$, the overall error probability is $\varepsilon_{(I)}$ and it holds $n=N/I$. Especially, $I=1$ indicates a one-shot transmission without ARQ/HARQ and $\varepsilon_{(0)}=0$. The monotonicity w.r.t. $n$ of equation \eqref{eq:error_HARQ}  implies that allowing retransmission in FBL-TDMA systems will only reduce the link reliability.



\section{Static Scheduling in Single-User Closed-Loop Communication}\label{sec:static_closedloop}

{In the problem discussed above}, the overall reward is the total successful transmission rate of all devices, where the priority of every individual transmission is the same and a fairness shall be achieved in the scheduling. Differing from that, we consider a closed communication loop between device and server, which is common in the emerging reliability-critical applications with closed-loop control such as automated factory and autonomous driving.

\subsection{Problem Setup}
For simplification, we consider one single device, for which the UL and DL transmissions share a fixed time frame {to fulfill the requirement of a guaranteed closed-loop air latency $T$}, and assume that ARQ can be executed in both directions with an extremely reliable ACK/NACK reliability and a minor feedback time cost $T\subscript{f}$. For both UL and DL, we consider all messages to have the same length of $d$ bits, and the channels to be block fading, i.e. both the dispersion and capacity remain consistent over a frame length $T$. \emph{In every individual frame, the DL transmission is enabled when and only when the UL transmission succeeds, and a unit reward is gained when and only when the DL transmission succeeds}. We look for an optimal strategy of retransmission scheduling and blocklength allocation in UL/DL that maximizes the expected reward $(1-\varepsilon\superscript{U})(1-\varepsilon\superscript{D})$ (the closed-loop reliability), where $\varepsilon\superscript{U}$ and $\varepsilon\superscript{D}$ are the message error rate in UL and DL, respectively.

\subsection{One-shot Scheme}
First, as a simple benchmark, we disable retransmissions and investigate the optimal UL/DL blocklength allocation. With no ARQ applied, the feedback cost $T\subscript{f}$ can be omitted. Similar to the TDMA case , we have the problem
\begin{maxi!}[2]
	{\left[n\superscript{U},n\superscript{D}\right]\in\mathbb{N}^2}{\left(1-\varepsilon\superscript{U}\right)\left(1-\varepsilon\superscript{D}\right)}{\label{prob:tdd_prob_one_shot}}{}
	\addConstraint{\left(n\superscript{U}+n\superscript{D}\right)T\subscript{S}\leqslant T\label{con:total_time}}
	\addConstraint{\max{\{\varepsilon\superscript{U},\varepsilon\superscript{D}\}}\le\varepsilon\subscript{max}.\label{con:max_err_rate}}
\end{maxi!}
and the packet error probabilities
\begin{align}
	\varepsilon\superscript{U}&\approx Q\left(\sqrt{\frac{n\superscript{U}}{V\superscript{U}}}(\mathcal{\mathcal{C}\superscript{U}}-\frac{d}{n\superscript{U}})\ln{2}\right),\label{eq:os_ul_err_rate}\\
	\varepsilon\superscript{D}&\approx Q\left(\sqrt{\frac{n\superscript{D}}{V\superscript{D}}}(\mathcal{\mathcal{C}\superscript{D}}-\frac{d}{n\superscript{D}})\ln{2}\right),\label{eq:os_dl_err_rate}
\end{align}
where $\mathcal{C}$ stands for the channel Shannon capacity, $V$ denotes the channel dispersion, $d$ is the message length, and $n$ stands for the dedicated  blocklength. The superscripts $(\cdot)\superscript{U}$ and $(\cdot)\superscript{D}$ denote the uplink and downlink, respectively, for distinction. \revise{Moreover, $\varepsilon\subscript{\max}$ is usually set to  no more than $0.5$, so that the constraint~\eqref{con:max_err_rate} forces the transmission rate to remain below the Shannon capacity, which represents the reliable transmission scenario. In case both constraints \eqref{con:total_time} and \eqref{con:max_err_rate} cannot be simultaneously fulfilled, Problem~\eqref{prob:tdd_prob_one_shot} is considered infeasible and no transmission shall be carried out.}

Similar to the classical case of FBL TDMA systems, due to the complexity optimization in integer space, we relax \eqref{prob:tdd_prob_one_shot} to $\left[n\superscript{U},n\superscript{D}\right]\in{\mathbb{R}^+}^2$, where we have the following theorem, for which the proof is given in Appendix~\ref{proof:one_shot_opt}:
\begin{theorem}\label{th:one_shot_opt}
	\revise{For any feasible resource scheme $\left[n\superscript{U}\subscript{os},n\superscript{D}\subscript{os}\right]$ of Problem~\eqref{prob:tdd_prob_one_shot}, the optimal one-shot allocation to maximize $\left(1-\varepsilon\superscript{U}\right)\left(1-\varepsilon\superscript{D}\right)$ is}
	obtained by solving
	\begin{equation}\label{eq:one_shot_opt}
	\begin{split}
		&\left(V\superscript{U}{\mathcal{C}\superscript{D}}^2+V\superscript{D}{\mathcal{C}\superscript{U}}^2\right){n\superscript{U}\subscript{os}}^3+\left(2dV\superscript{U}\mathcal{C}\superscript{D}-2n_\Sigma V\superscript{U}{\mathcal{C}\superscript{D}}^2\right.\\
		&\left.-n_\Sigma V\superscript{D}{\mathcal{C}\superscript{U}}^2-2dV\superscript{D}\mathcal{C}\superscript{U}\right){n\superscript{U}\subscript{os}}^2+\left(n_\Sigma^2V\superscript{U}{\mathcal{C}\superscript{D}}^2+d^2V\superscript{U}\right.\\
		&\left.-2dn_\Sigma V\superscript{U}\mathcal{C}\superscript{D}+2dn_\Sigma V\superscript{D}\mathcal{C}\superscript{U}+d^2V\superscript{D}\right)n\superscript{U}\subscript{os}-d^2n_\Sigma V\superscript{D}=0
	\end{split}
	\end{equation}
	and  $n\superscript{D}\subscript{os}=n_\Sigma-n\superscript{U}\subscript{os}$, where $n_\Sigma\triangleq\frac{T}{T\subscript{S}}$.
\end{theorem}
Remark that in the special case of TDD scenarios, where the UL and DL share the same radio channel and are therefore symmetric, i.e. $V\superscript{U}=V\superscript{D}$ and $\mathcal{C}\superscript{U}=\mathcal{C}\superscript{D}$, Eq.~\eqref{eq:one_shot_opt} returns an unique root $n\superscript{U}\subscript{os}=n\superscript{D}\subscript{os}=\frac{n_\Sigma}{2}$.

\subsection{Static Scheduling with HARQ}\label{subsec:static_HARQ_scheduling}
Then we investigate the performance of HARQ under a static retransmission scheduling, where the retransmission times $I$ and the blocklength allocation among slots are prefixed before the transmission. \revise{Consider the time frame $T$ fully utilized by $I$ UL slots $\mathbf{t}\superscript{U}=\left[n\superscript{U}_1T\subscript{S}, n\superscript{U}_2T\subscript{S}\dots n\superscript{U}_IT\subscript{S}\right]$ and $I$ DL slots $\mathbf{t}\superscript{D}=\left[n\superscript{D}_1T\subscript{S}, n\superscript{D}_2T\subscript{S}\dots n\superscript{D}_IT\subscript{S}\right]$, and define $\mu\superscript{loop}_{(I), \text{s}}$ as the corresponding closed-loop reliability, we propose the following optimization problem:
\begin{maxi!}[2]
	{I,\mathbf{t}\superscript{U},\mathbf{t}\superscript{D}}{\mu\superscript{loop}_{(I), \text{s}}}{}{}
	\addConstraint{\sum\limits_{i=1}^{I}\left(n\superscript{U}_iT\subscript{S}+n\superscript{D}_iT\subscript{S}\right)+(2I-1)T\subscript{f}=T}
	\addConstraint{\varepsilon_{(i)}\superscript{U}\leqslant\varepsilon\subscript{max},\varepsilon\superscript{D}_{(i)}\leqslant\varepsilon\subscript{max},\forall i\in\{1,2,\dots I\},}
\end{maxi!}
where $\varepsilon_{(i)}\superscript{U}$ and $\varepsilon_{(i)}\superscript{D}$ are the lower bound of error probabilities up to the $i\superscript{th}$ transmission attempt with ideal incremental redundancy HARQ in UL and DL, respectively.} We propose the following lemma with proof in Appendix~\ref{proof:static_retr_opt}:
\begin{lemma}\label{lem:static_retr_opt}
Under static retransmission scheduling, the closed-loop reliability is maximized by the optimal one-shot scheme, i.e., $I=1$, $n\superscript{U}=n\superscript{U}\subscript{os}$, and $n\superscript{D}=n\superscript{D}\subscript{os}$.
\end{lemma}
The integer {root} can be then approximated by rounding the real-valued one-shot optimum:
\begin{equation}
		\arg\max\limits_{[I, n\superscript{U}, n\superscript{D}]\in\mathbb{N}^3}\mu\superscript{loop}_{(I), \text{s}}\approx\left[1,\left\lceil n\superscript{U}\subscript{os}\right\rfloor, \left\lceil n\superscript{D}\subscript{os}\right\rfloor\right].
\end{equation}

\section{Dynamic Retransmission Protocol}\label{sec:clarq}
\subsection{Protocol Design}
So far, we have shown that the one-shot scheme towards equal transmission error probabilities in UL and DL is the optimum among all static retransmission schedules. Now we consider a dynamic retransmission scheme, where the device and the server are able to reschedule the blocklength allocation upon the ACK/NACK feedback for the last transmission attempt.

We begin with the initial scheduling before the first UL attempt. As shown by \eqref{eq:ul_err_rate_bound}, with any arbitrary time $t\superscript{U}_1$ scheduled for UL at this stage, the optimal solution is always to use it entirely for a one-shot UL transmission attempt.

Then investigate the rational decision of the device after making an unsuccessful UL transmission attempt. Knowing about the last message error in UL from the NACK feedback, the device should always schedule another UL transmission attempt with the remaining time, which was previously reserved for the DL, until the remaining time falls below a certain threshold to ensure a minimal chance of successful transmission loop. This policy is self-evidently transmission-rate-optimal, as the reward will always be $0$ if the device fails to transmit its message in UL.

On the other hand, once the device succeeds {in an} UL attempt and obtains an ACK feedback from the server, any further retransmission in UL will certainly bring no extra reward, but only waste the time resource. Hence, the remaining time should be completely exploited for the DL transmission. Furthermore, as proven in Appendix {\ref{proof:static_retr_opt}}, given a fixed amount of remaining time, the optimum is to entirely exploit it for a one-shot DL transmission.

\begin{figure}[!tb]
	\centering
	\includegraphics[width=.65\linewidth]{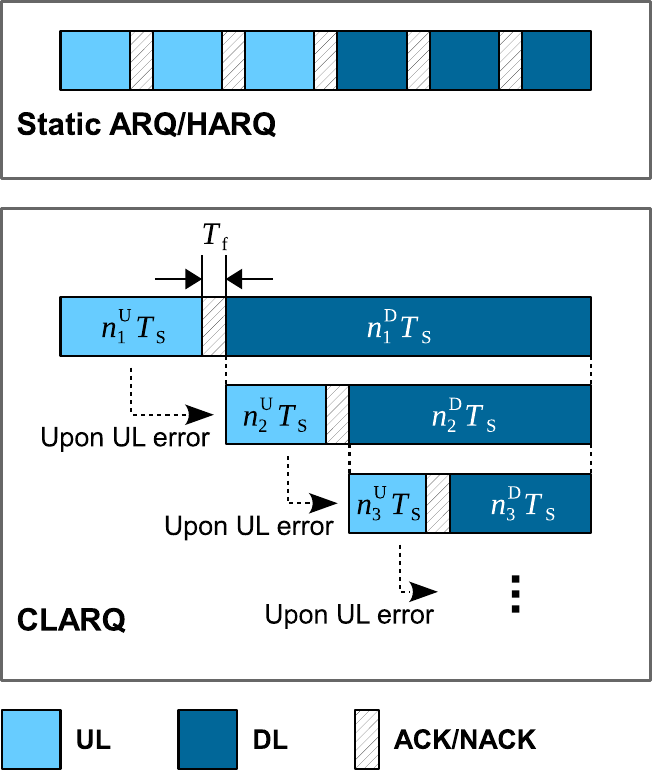}
	\caption{Dynamic blocklength allocation with the CLARQ protocol in comparison to the static ARQ/HARQ scheduling as described in Section~\ref{subsec:static_HARQ_scheduling}.}
	\label{fig:clarq}
\end{figure}

Thus, we propose the protocol of Closed-Loop ARQ (CLARQ),  {which is described by Algorithm~\ref{alg:clarq}. With CLARQ, a device recursively re-allocates the remaining blocklength in current time frame for the next UL attempt, until it exceeds the limit for retransmissions, or receives an ACK for UL success and thereby assigns all remaining blocklength to one DL slot. In this approach, both the blocklength assigned to DL slot and therefore the DL error rate are dynamically determined by the UL results. In contrast, under static ARQ/HARQ with fixed frame length, the blocklength is pre-determined for all transmission slots, so the error rates in UL and DL are independent from each other, as illustrated in Fig.~\ref{fig:clarq}.}

\begin{alg}{The CLARQ Protocol}{}{}
	\removelatexerror
	\begin{algorithm}[H]
		\label{alg:clarq}
		\footnotesize
		\DontPrintSemicolon
		Initialization: $i=0$, $T\superscript{U}\subscript{min}$, $T\superscript{D}\subscript{min}$, $n\superscript{D}_0=T/T\subscript{S}$, $\text{ACK}=\text{false}$\;
		\While(\hfill\emph{Retransmit in UL until a success}){$\text{ACK}$}{
			\uIf(\hfill\emph{(Re)attempt}){$n\superscript{D}_i\ge(T\superscript{U}\subscript{min}+T\superscript{D}\subscript{min}+T\subscript{f})/T\subscript{S}$}{
				$i=i+1$\;
				Reschedule $\left(n\superscript{U}_i+n\superscript{D}_i\right)T\subscript{S}+T\subscript{f}\leqslant T\superscript{D}_{i-1}$\;
				Retransmit in UL with $n\superscript{U}_iT\subscript{S}$\;
				Update ACK\;
			}
			\Else(\hfill\emph{Insufficient time remaining}){Break\;}
		}
		Transmit in DL with $n\superscript{D}_iT\subscript{S}$\;
	\end{algorithm}
\end{alg}

The minimal transmission slot lengths in UL and DL, namely $T\superscript{U}\subscript{min}$ and $T\superscript{D}\subscript{min}$, are set due to the concern that when the blocklength falls below some lower bound, FBL's error rate dramatically increases to an unacceptable level, where any transmission attempt will hardly succeed. They are commonly selected according to an maximal packet error rate $\varepsilon\subscript{max}$:
\begin{align}
		T\superscript{U}\subscript{min}&=T\subscript{S}\times\arg\left(n\superscript{U}\vert\varepsilon\superscript{U}=\varepsilon\subscript{max}\right)\nonumber\\&=T\subscript{S}\times\frac{\sqrt{\beta\superscript{U}(\varepsilon\subscript{max})^2+4\mathcal{C}\superscript{U}d}-\beta\superscript{U}(\varepsilon\subscript{max})}{2\mathcal{C}\superscript{U}}\\
		T\superscript{D}\subscript{min}&=T\subscript{S}\times\arg\left(n\superscript{D}\vert\varepsilon\superscript{D}=\varepsilon\subscript{max}\right)\nonumber\\&T\subscript{S}\times\frac{\sqrt{\beta\superscript{D}(\varepsilon\subscript{max})^2+4\mathcal{C}\superscript{D}d}-\beta\superscript{D}(\varepsilon\subscript{max})}{2\mathcal{C}\superscript{D}}
\end{align}
where $\beta\superscript{U}(\varepsilon)=-\frac{\sqrt{2V\superscript{U}}\text{erfc}^{-1}(2\varepsilon)}{\ln 2}$ and $\beta\superscript{D}(\varepsilon)=-\frac{\sqrt{2V\superscript{D}}\text{erfc}^{-1}(2\varepsilon)}{\ln 2}$. The inverse cumulative error function $\text{erfc}^{-1}(\cdot)$ has no closed analytical form, yet it can be conveniently approximated with sufficient accuracy \cite{Borjesson1979simple}.

At the beginning of every iteration of the \emph{while}-loop in Algorithm \ref{alg:clarq}, there has to be at least a time of $T\superscript{U}\subscript{min}+T\superscript{D}\subscript{min}+T\subscript{f}$ remaining, in order to support a new UL attempt its corresponding DL transmission, as examined with the \emph{if}-condition. This also implies that, denoted by $n\superscript{U}_i$ and $n\superscript{D}_i$ the UL and DL blocklength in the $i\superscript{th}$ (re)schedule, respectively, an arbitrary feasible schedule $\mathbf{n}_I=\left[n\superscript{U}_1, n\superscript{D}_1, n\superscript{U}_2, n\superscript{D}_2\dots n\superscript{U}_I, n\superscript{D}_I\right]$ with up to $I$ UL attempts must fulfill
\begin{equation}
	n\superscript{D}_{i-1}T\subscript{S}\ge \left(n\superscript{U}_{i}+n\superscript{D}_{i}\right)T\subscript{S}+T\subscript{f},\quad\forall i\in\{1, 2\dots I\},
\end{equation}
where for $i=0$ it is defined $\varepsilon_0\superscript{U}=1$ and $n\superscript{D}_0T\subscript{S}=T$.

Furthermore, the closed-loop reliability of this schedule is
\begin{equation}
	\mu\superscript{loop}_{(I),\text{d}}=\sum\limits_{i=1}^{I}\left\{\left[\left(1-\varepsilon\superscript{D}_i\right)\left(1-\varepsilon\superscript{U}_i\right)\right]\prod\limits_{j=0}^{i-1}\varepsilon\superscript{U}_j\right\}
	\label{eq:closed_loop_reward}
\end{equation}

\subsection{CLARQ Optimization in Bellman's View}\label{subsec:bellman_view}
\revise{While Algorithm~\ref{alg:clarq} is only outlining a protocol without any performance control, now we consider its optimization regarding $\mu\superscript{loop}_{(I),\text{d}}$}:

\begin{maxi!}[2]
	{I,\mathbf{n}_I}{\sum\limits_{i=1}^{I}&\left\{\left[\left(1-\varepsilon\superscript{D}_i\right)\left(1-\varepsilon\superscript{U}_i\right)\right]\prod\limits_{j=0}^{i-1}\varepsilon\superscript{U}_j\right\}}{\label{eq:tdd_prob}}{}
	\addConstraint{\left(n\superscript{U}_i+n\superscript{D}_i\right)T\subscript{S}+T\subscript{f}\leqslant n\superscript{D}_{i-1}T\subscript{S}\label{eq:reallocation_constraint_with_tf}}
	\addConstraint{n\superscript{U}_{i}T\subscript{S}\ge T\superscript{U}\subscript{min},\quad n\superscript{D}_{i}T\subscript{S}\ge T\superscript{D}\subscript{min}.\label{eq:tmin_constraint}}
\end{maxi!}

Following the common approach of analyzing ARQ/HARQ performance in the FBL regime, {which is widely applied in literature such as \cite{Avranas2018energy} and \cite{Makki2019fast},} here we consider a negligible feedback time loss $T\subscript{f}\approx 0$\footnote{For TDD systems, this approximation can widely hold in most practical scenarios,  and greatly improve the convenience of analysis to the upper bound of HARQ performance. For FDD systems, it is even technically capable to implement the system in a way that $T\subscript{f}=0$, e.g. by embedding the ACK/NACK into the first bits of the DL message.}. Thus, the constraint \eqref{eq:reallocation_constraint_with_tf} becomes
\begin{equation}
	\forall i\in\{1, 2\dots I\}:\quad n\superscript{U}_i+n\superscript{D}_i\leqslant n\superscript{D}_{i-1};\label{eq:reallocation_constraint_without_tf}
\end{equation}
With a certain $I$, the optimization of multi-stage allocation \eqref{eq:tdd_prob} is a $2I$-dimensional integer programming problem with $2I$ linear constraints set by \eqref{eq:tmin_constraint} and \eqref{eq:reallocation_constraint_without_tf}. Such problems are known to be NP-Hard. To make it {worse}, the optimal $I$ is unknown in our problem.

Following the classical FBL information theoretic approaches, for this moment we relax the space of $\mathbf{n}_I$ from ${\mathbb{N}^+}^{2I}$ to ${\mathbb{R}^+}^{2I}$, and remove the constraint \eqref{eq:tmin_constraint}. In this case, as $\varepsilon\superscript{U}_i$ and $\varepsilon\superscript{D}_i$ monotonically decrease w.r.t. $n_i\superscript{U}$ and $n_i\superscript{D}$, respectively, it is trivial to prove that the maximum, if any, must fulfill
\begin{equation}
	n_i\superscript{U}+n_i\superscript{D}=n_{i-1}\superscript{D}, \forall i\in\{1,2\dots I\}.\label{eq:blocklength_iteration}
\end{equation}
Thus, $\mathbf{n}_I$ can be uniquely determined by its sub-sequence $\tilde{\mathbf{n}}_I=\left[n\superscript{U}_1,n\superscript{U}_2\dots n\superscript{U}_I, n\superscript{D}_I\right]$, and we can exploit the recursive feature of \eqref{eq:tdd_prob} under constraint \eqref{eq:reallocation_constraint_without_tf} that
\begin{equation}{
	\footnotesize
	\begin{split}
		&\max_{\vert\tilde{\mathbf{n}}_I\vert\le\frac{T}{T\subscript{S}}}\mu\superscript{loop}_{(I),\text{d}}=\max_{\vert\tilde{\mathbf{n}}_I\vert= \frac{T}{T\subscript{S}}}\mu\superscript{loop}_{(I),\text{d}}\\
		=&\max_{0\leqslant n_1\superscript{U}\le\frac{T}{T\subscript{S}}}\left\{(1-\varepsilon\superscript{D}_1)(1-\varepsilon\superscript{U}_1)+\varepsilon\superscript{U}_1\max_{\vert\tilde{\mathbf{n}}_{I}\vert=\frac{T}{T\subscript{S}}-n_1\superscript{U}}\mu\superscript{loop}_{(I),\text{d}}\right\},
		\label{eq:bellman_eq}
	\end{split}
}
\end{equation}
which decomposes the original problem into $I$ single-stage problems, where the condition of $(i+1)\superscript{th}$ stage is uniquely fixed by the result of the $i\superscript{th}$ stage. This is known as the dynamic programming (DP) approach, where \eqref{eq:bellman_eq} is called the Bellman equation. Its optimum is uniquely determined by the PER functions $\varepsilon_i\superscript{U}$ and $\varepsilon_i\superscript{D}$ of all individual stages $i$. While $\varepsilon_i\superscript{U}$ and $\varepsilon_i\superscript{D}$ highly depend on the encoding and combining performance of the specific HARQ scheme, in this paper we analyze the case of simple ARQ without information combining, which is a tight lower performance bound of all HARQ schemes

With simple ARQ, for all $i\in\{1, 2\dots I\}$, we have in UL $\varepsilon_i\superscript{U}\approx Q\left(\sqrt{\frac{n_i\superscript{U}}{V}}(\mathcal{C}\superscript{U}-r_i)\ln{2}\right)$ and in DL $\varepsilon_i\superscript{D}\approx Q\left(\sqrt{\frac{n_i\superscript{D}}{V}}(\mathcal{C}\superscript{D}-r_i)\ln{2}\right)$. In this case we can provide the following theorem and corollaries, as proved in Appendices~\ref{proof:concavity} and \ref{proof:uniq_solution}, respectively.

\begin{lemma}\label{lem:concavity}
	The success rate of the $i^{\rm th}$ round transmission attempt $\mu_{i,\text{d}}\superscript{loop}=(1-\varepsilon_i\superscript{U})(1-\varepsilon_i\superscript{D})$ is concave w.r.t. $n_i\superscript{U}$ over $\left[n\subscript{min}\superscript{U}, n_{i-1}\superscript{D}\right]$, where $n\superscript{U}\subscript{min}=\frac{T\superscript{U}\subscript{min}}{T\subscript{S}}$.
\end{lemma}

\begin{theorem}\label{th:uniq_solution}
	With sufficient $T$, the Bellman equation \eqref{eq:bellman_eq} has a unique solution $\tilde{\mathbf{n}}_{I,\text{opt}}$, which fulfills $n\superscript{U}_{1,\text{opt}}\ge n\superscript{U}_{2,\text{opt}}\ge\dots\ge n\superscript{U}_{I,\text{opt}}$ and $\varepsilon_I\superscript{U}\left(n\superscript{U}_{I,\text{opt}}\right)=\varepsilon_I\superscript{D}\left(n\superscript{D}_{I,\text{opt}}\right)$.
\end{theorem}

Especially, for TDD systems where the channel is symmetric in UL and DL, we have the following corollaries, for which the proofs are provided in Appendices~\ref{proof:upper_lower_bounds_tdd} and \ref{proof:number_stages}, respectively.

\begin{corollary}\label{cor:upper_lower_bounds_tdd}
	With simple ARQ, the optimal schedule $\tilde{\mathbf{n}}_{I,\text{opt}}$ of a TDD system always guarantees $n\superscript{U}_{i,\text{opt}}\in\left[n\subscript{min}, 2^{I-i+1}n\subscript{min}\right)$ and $n\superscript{D}_{i,\text{opt}}\in\left[(I-i+1)n\subscript{min}, 2^{I-i+1}n\subscript{min}\right)$ for all $i\in\{1,2\dots I\}$, where $n\subscript{min}=n\superscript{U}\subscript{min}=n\superscript{D}\subscript{min}$.
\end{corollary}

\begin{corollary}\label{cor:number_stages}
	With simple ARQ, the maximal number of UL transmission attempts $I$ in the optimal TDD schedule $\tilde{\mathbf{n}}_{I,\text{opt}}$ is bounded in the interval $\left(\log_2\left(\frac{T}{T\subscript{min}}\right)-1, \frac{T}{T\subscript{min}}-1\right]$
\end{corollary}

\subsection{CLARQ Optimization through Integer DP}
To solve problems {like} \eqref{eq:bellman_eq}, it {generally} needs to define the reward function of a single step action (allocation). In the case of CLARQ, denote $\theta_i(n_i\superscript{U})$ the reward of $i\superscript{th}$ (re-)transmission, which is the sum of expected closed-loop reliability from the $i\superscript{th}$ to the last attempt:
\begin{equation}
\begin{split}
&\theta_i(n_{i,\text{opt}}\superscript{U})
=\max\limits_{n\superscript{U}_i}\left[\mu_{i,\text{d}}\superscript{loop}+\varepsilon\superscript{U}_i \theta_{i+1}\right]\\
=&\max\limits_{n\superscript{U}_i}\left[(\theta_{i+1}-1)\varepsilon_i\superscript{U}+1-\varepsilon\superscript{D}_i+\varepsilon\superscript{U}_i\varepsilon\superscript{D}_i\right]
\end{split}
\label{eq:recursion_bellman}
\end{equation}
We can therewith recursively solve $n_i\superscript{U}$ with $\theta_{i+1}$, backwards from $i=I$ to $i=1$. Especially, noting that $\theta_{I+1}=0$, so we have $\theta_I=(1-\varepsilon\superscript{U}_I)^2$ at  $\varepsilon_{I,\text{opt}}\superscript{U}=\varepsilon_{I,\text{opt}}\superscript{D}$ as Theorem \ref{th:uniq_solution} suggests. Yet the (global) optimal value $n_{I,\text{opt}}\superscript{U}$ is unknown, but its upper and lower bounds are provided by Corollary~\ref{cor:upper_lower_bounds_tdd}, so we search for the optimum by testing different values over the solution space, and for each specific value of $n_I\superscript{U}$ we can obtain a sequence $\left[n_{I-1}\superscript{U}, n_{I-2}\superscript{U}\dots n_1\superscript{U}\right]$ by recursively solving \eqref{eq:theta_extreme} for $i=1,2\dots I-1$.

For the relaxed problem in real-vector space, another challenge is met here that \eqref{eq:theta_extreme} has analytical solution only when $\theta_{i + 1}=0$, i.e. for the last stage $i=I$. For early stages $i\in\{1,2\dots i-1\}$, it has to rely on numerical methods to approximate the optimum $n_{i,\text{opt}}\superscript{U}$ in the infinite space $\mathbb{R}^+$.

\begin{strip}
\begin{alg}{Integer CLARQ Policy Optimization}{}{}
	\removelatexerror
	\begin{algorithm}[H]
		\label{alg:int_clarq_opt}
		\scriptsize
		\DontPrintSemicolon
		\SetKwFunction{FMain}{Main}
		\SetKwFunction{FReward}{Rwd}
		\SetKwFunction{FBestReward}{BestRwd!}
		\SetKwProg{Pn}{Function}{:}{}
		\Pn{\FMain{$T\subscript{S}, T\superscript{U}\subscript{min}, T\superscript{D}\subscript{min}$}}
		{
			$n\subscript{max}=\left\lfloor\frac{T}{T\subscript{S}}\right\rfloor, n\subscript{min}\superscript{U}=\left\lceil\frac{T\superscript{U}\subscript{min}}{T\subscript{S}}\right\rceil,n\subscript{min}\superscript{D}=\left\lceil\frac{T\superscript{D}\subscript{min}}{T\subscript{S}}\right\rceil$\;
			$\mathbf{\Phi}\gets\mathbf{0}_{n\subscript{max}\times1}, \mathbf{\Xi}\gets\mathbf{0}_{n\subscript{max}\times1}$\;
			 \FBestReward{$n,\mathbf{\Phi},\mathbf{\Xi},n\subscript{min}\superscript{U},n\subscript{min}\superscript{D}$}\hspace{9.5cm}\hfill\emph{\raggedleft Calculate the optimum}\;
			$n=n\subscript{max}, i=0$\;
			\While(\hfill\emph{Read the allocation stage-by-stage}){$n\ge n\subscript{min}\superscript{U}+n\subscript{min}\superscript{D}$}{
				$n_{i,\text{opt}}\superscript{U}=\phi_{n}$\;
				$n= n-\phi_{n}, i=i+1$\;
			}
			$I= i, n_{I,\text{opt}}\superscript{D}= n_{I,\text{opt}}\superscript{U}$\;
		}
		\Return $\left[n_{1,\text{opt}}\superscript{U}, n_{2,\text{opt}}\superscript{U}\dots n_{I,\text{opt}}\superscript{U}, n_{I,\text{opt}}\superscript{D}\right]$\;
		\;
		\Pn{\FBestReward{$n,\mathbf{\Phi},\mathbf{\Xi},n\subscript{min}\superscript{U},n\subscript{min}\superscript{D}$}}{
			\uIf(\hfill\emph{Pre-calaulated}){$\phi_n>0$}{
				\Return $\xi_n$\;
			}
			\uElseIf(\hfill\emph{Insufficient for UL+DL}){$n< n\subscript{min}\superscript{U}+n\subscript{min}\superscript{D}$}{
				$\phi_n\gets n$\;
				$\xi_n\gets 0$\;
			}
			\uElseIf(\hfill\emph{{Identify the $I\superscript{th}$ stage}}){$n< 2n\subscript{min}\superscript{U}+n\subscript{min}\superscript{D}$}{
				$\phi_n\gets \arg\max\limits_{m}\left\{\FReward\left(n,m,0\right)\right\}$\;
				$\xi_n\gets\FReward\left(n,\phi_n, 0\right)$\;
			}
			\Else(\hfill\emph{Recursion}){
				$\phi_n\gets \arg\max\limits_{m}\left\{\FReward\left(n,m, \FBestReward\left(n-m,\mathbf{\Phi},\mathbf{\Xi},n\subscript{min}\superscript{U},n\subscript{min}\superscript{D}\right)\right)\right\}$\;
				$\xi_n\gets\FReward\left(n,\phi_n, \FBestReward\left(n-\phi_n,\mathbf{\Phi},\mathbf{\Xi},n\subscript{min}\superscript{U},n\subscript{min}\superscript{D}\right)\right)$\;
			}
		}\Return $\xi_n$\;
		\;
		\Pn{\FReward{$n,m,\theta\subscript{f}$}}{}
		\Return $\left[1-\varepsilon(m)\right]\left[1-\varepsilon(n-m)\right]+\varepsilon(m)\theta\subscript{f}$\;
	\end{algorithm}
\end{alg}
\end{strip}

Nevertheless, remark that the real-value relax was taken, like in classical FBL works, only for the convenience of analysis, while the final solution ${\mathbf{n}}_{I,\text{opt}}$ of \eqref{eq:tdd_prob} can only take values in ${\mathbb{N}^+}^{2I}$, which is a finite integer vector space under the boundaries provided by Corollaries~\ref{cor:upper_lower_bounds_tdd} and \ref{cor:number_stages}. Furthermore, to apply FBL approaches, $\frac{T}{T\subscript{min}}$ has to be -- as referred earlier -- strictly limited, i.e. the solution space is usually of a reasonable size. This enables to apply classical dynamic programming (DP) techniques to directly solve the integer global optimum, which is typically realized through a recursive computation algorithm accompanied with memory over the solution space, as described 
{by Algorithm~\ref{alg:int_clarq_opt}:}
\begin{itemize}
	\item Two global vectors are defined, namely $\mathbf{\Phi}=[\phi_1,\phi_2\dots\phi_{n\subscript{max}}]$ and $\mathbf{\Xi}=[\xi_1,\xi_2\dots \xi_{n\subscript{max}}]$, in order to store $n_{i,\text{opt}}\superscript{U}$ as function of $n_{i-1}\superscript{D}\in[1,n\subscript{max}]$, and the corresponding rewards $\theta_i(n_{i,\text{opt}\superscript{U}})$, respectively.
	\item The function \FBestReward is implemented to recursively solve $n_{i,\text{opt}}\superscript{U}$ and $\theta_i\left(n_{i,\text{opt}\superscript{U}}\right)$, and \emph{therewith update the global variables} $\mathbf{\Phi}$ and $\mathbf{\Xi}$, respectively.
	\item The function \FReward is called by \FBestReward to calculate the reward of an arbitrary given blocklength allocation with known future reward.
	\item The \FMain function calls \FBestReward to solve the problem, and returns the allocation in a structured format.
\end{itemize}

\subsection{Computational Complexity Analysis}\label{subsec:computation_complexity}
{In the last subsection, Algorithm~\ref{alg:int_clarq_opt} implements a recursive DP approach}, i.e., the \FBestReward function, to compute the integer dynamic program. The recursive algorithm considers the CLARQ as an ${O(n\subscript{max})}$ stage DP problem with the vector ${\mathbf{\Phi}}$ used as the DP memory. In the recursive process, each of its ${n}$-stage sub CLARQ problem, where ${n<n\subscript{max}}$, should be solved and stored in the element ${\phi_n}$ of ${\mathbf{\Phi}}$, after and only after: 
\begin{enumerate}
	\item The ${n}$-stage sub CLARQ problem is first-time called by the \FBestReward function;
	\item For the given ${n}$-stage sub CLARQ problem, each of its own ${m}$-stage sub CLARQ problems, where ${m<n}$, is already solved and stored in ${\phi_m}$, respectively. 
\end{enumerate}

It is straight forward that the space complexity of Algorithm 2 is given by the length of the DP memory ${\mathbf{\Phi}}$, i.e., ${O(n\subscript{max})}$. To analyze the algorithm's time complexity, note that it only takes ${O(n)}$ time to solve and store each ${n}$-stage sub CLARQ problem if every ${m}$-stage sub CLARQ problem of the ${n}$-stage problem is already solved and stored in ${\mathbf{\Phi}}$. Since the time complexity of running the \FBestReward function for a ${n\subscript{max}}$-stage problem is bounded by the total time complexity to solve-and-store all the $i\superscript{th}$-stage sub CLARQ problems iteratively, i.e., to solve them in the ascending order ${i=1,2\dots n\subscript{max}}$, we can conclude that the total time complexity of the recursive DP algorithm is ${O(n\subscript{max}^2)}$.

{In realistic scenarios of deployment, such a time complexity can critically challenge the online computation of optimal CLARQ policy regarding the time-varying channel condition, leading to a violation of the real-time performance and a short battery life of mobile devices. To address this issue, it is a practical solution to rely on a look-up-table (LUT) that contains a set optimal CLARQ policies in various channel conditions, which were offline solved in priori and programmed into the devices. Thus, the devices can be rapidly adapted to the appropriate specification w.r.t. the real-time channel measurement, with only a minimal time complexity as low as ${O(I)}$. Additionally, regarding the offline computation, analytical results from Lemma 1 and addtional scheduling bounds from Corollary 2 are also implemented in the \FBestReward function, in order to avoid solving unnecessary sub DP problems and accelerate the algorithm's computational efficiency.}

\section{Numerical Simulations}\label{sec:eval}
{Extensive numerical simulation campaigns are carried out by means of a mathematical commercial tool, namely MATLAB, where our novel approach CLARQ is developed and optimization policies are properly executed.}

\subsection{CLARQ Policy and Performance Analysis}
To demonstrate our proposed CLARQ protocol and the DP optimizing method, we set the following system specifications as listed in Table~\ref{tab:sys_spec}. Two different scenarios A and B are defined as samples of symmetric and asymmetric UL/DL channels, respectively. 
\begin{table}[!htb]
	\centering
	\caption{System specifications for evaluation}
	\label{tab:sys_spec}
	\begin{tabular}{cccl}
		\toprule[2px]
		\textbf{Parameter} & \multicolumn{2}{c}{\textbf{Value}} & \textbf{Description}\\
		\midrule[1.5px]
		Modulation scheme & \multicolumn{2}{c}{BPSK} &\\\hline
		$T\subscript{S}$ & \multicolumn{2}{c}{\SI{4}{\micro\second}} & Symbol length\\\hline
		$T$ & \multicolumn{2}{c}{\SI{10}{\milli\second}} & Maximal closed-loop air latency\\\hline
		$n\subscript{max}$ & \multicolumn{2}{c}{2500 bit}  & Total blocklength available\\\hline
		$d$ & \multicolumn{2}{c}{16 bit} & Uncoded packet size\\\hline
		$\varepsilon\subscript{max}$ & \multicolumn{2}{c}{$0.2$} & Maximal packet error rate\\\hline
		\multirow{2}{*}{$\gamma\superscript{U}$} & A: & \SI{-13}{\dB} & \multirow{2}{*}{SNR in UL}\\
		& B: & \SI{-11}{\dB} &\\\hline
		\multirow{2}{*}{$\gamma\superscript{D}$} & A: & \SI{-13}{\dB} & \multirow{2}{*}{SNR in DL}\\
		& B: & \SI{-15}{\dB} &\\\hline
		\multirow{2}{*}{$n\subscript{min}\superscript{U}$} & A: & 322 bit & \multirow{2}{*}{Minimal blocklength in UL}\\
		& B: & 232 bit &\\\hline
		\multirow{2}{*}{$n\subscript{min}\superscript{D}$} & A: & 322 bit & \multirow{2}{*}{Minimal blocklength in DL}\\
		& B: & 533 bit &\\
		\bottomrule[2px]
	\end{tabular}
\end{table}

\begin{figure*}[!htbp]
	\centering
	\begin{subfigure}[t]{0.48\linewidth}
		\centering
		\includegraphics[width=.7\linewidth]{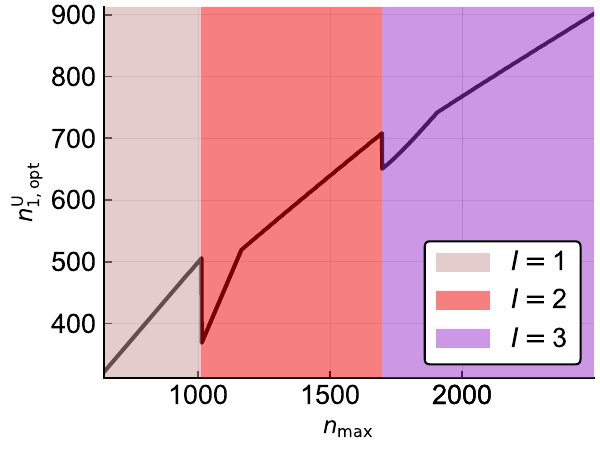}
		\caption{First UL slot, scenario A}
		\label{subfig:demo_first_ul_blocklength_sym}
	\end{subfigure}
	\begin{subfigure}[t]{0.48\linewidth}
		\centering
		\includegraphics[width=.7\linewidth]{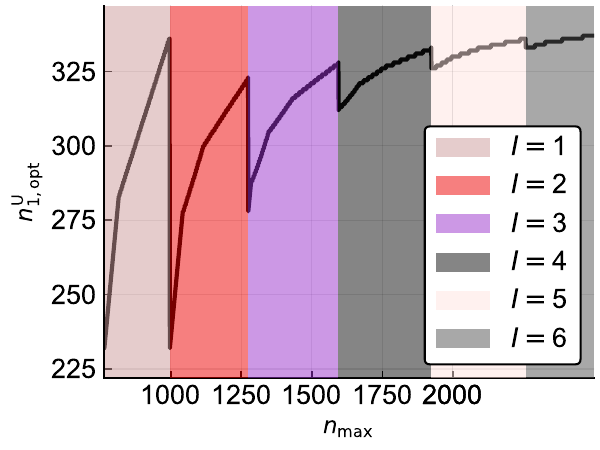}
		\caption{First UL slot, scenario B}
		\label{subfig:demo_first_ul_blocklength_asym}
	\end{subfigure}
	\begin{subfigure}[t]{0.48\linewidth}
		\centering
		\includegraphics[width=.7\linewidth]{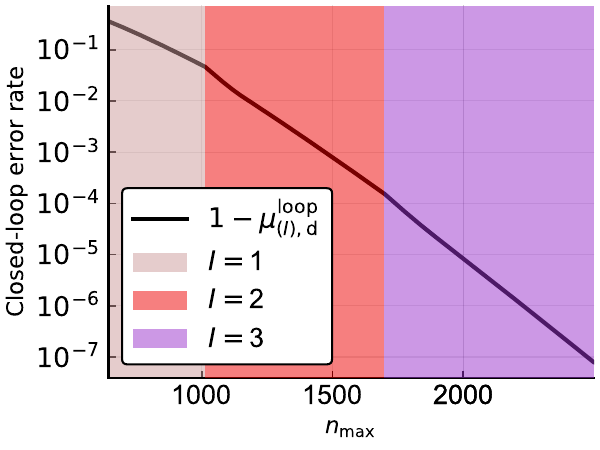}
		\caption{Closed-loop error rate, scenario A}
		\label{subfig:demo_closed_loop_err_sym}
	\end{subfigure}
	\begin{subfigure}[t]{0.48\linewidth}
		\centering
		\includegraphics[width=.7\linewidth]{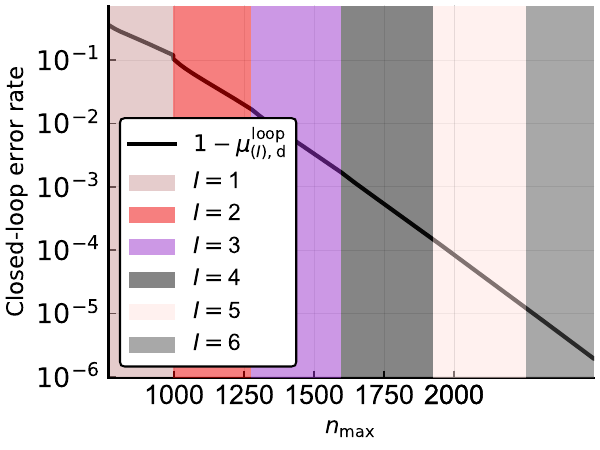}
		\caption{Closed-loop error rate, scenario B}
		\label{subfig:demo_closed_loop_err_asym}
	\end{subfigure}
	\begin{subfigure}[t]{0.48\linewidth}
		\centering
		\includegraphics[width=.8\linewidth]{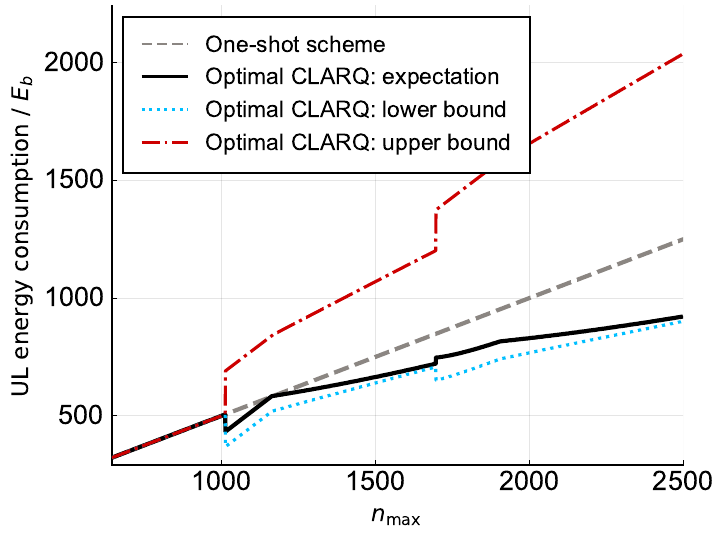}
		\caption{{Power consumption, scenario A}}
		\label{subfig:demo_power_efficiency_sym}
	\end{subfigure}
	\begin{subfigure}[t]{0.48\linewidth}
		\centering
		\includegraphics[width=.8\linewidth]{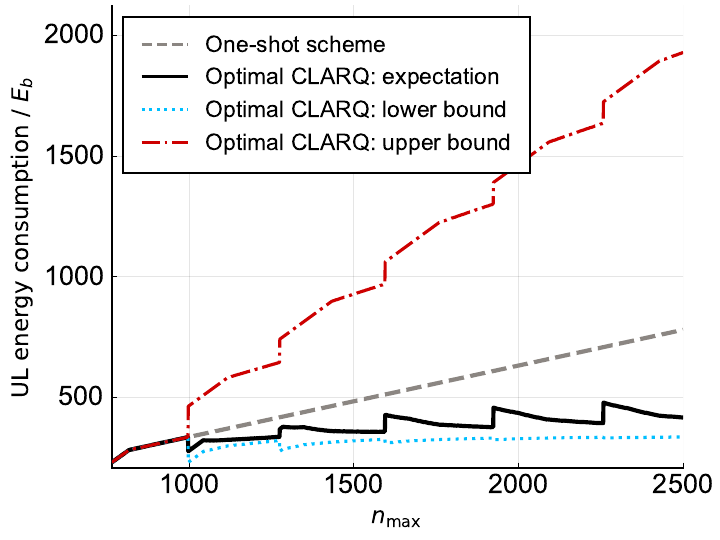}
		\caption{{Power consumption, scenario B}}
		\label{subfig:demo_power_efficiency_asym}
	\end{subfigure}
	\caption{{The blocklength of first UL slot $n_{1,\text{opt}}\superscript{U}$ as function of $n\subscript{max}$, which sufficiently presents the optimal CLARQ policy; the corresponding closed-loop error rate; and the uplink energy consumption per frame $T$, which dominates the power efficiency.}}\label{fig:demo}
\end{figure*}

To better understand the behavior of optimal CLARQ policy, we investigate the first UL slot length $n_{i,\text{opt}}\superscript{U}$ and the maximal times $I$ of UL (re)transmission attempts for different values of $n\subscript{max}$, as illustrated in Figs.~\ref{subfig:demo_first_ul_blocklength_sym} and ~\ref{subfig:demo_first_ul_blocklength_asym}. We can observe that $n_{1,\text{opt}}\superscript{U}$ is a segmented function of $n\subscript{max}$, which is monotonically increasing in its every individual segment. The discontinuity of the function roots in the dynamic increase of maximal retransmission attempts $I$. Since the Bellman Equation~\eqref{eq:bellman_eq} is consistent to $I$ and $T$ (i.e. $n\subscript{max}$), this curve actually describes the complete optimal CLARQ policy under the given system specification. For instance, in scenario A where $n\subscript{min}\superscript{U}=n\subscript{min}\superscript{D}=322$, starting with $n_{i,\text{opt}}\superscript{U}\vert_{n\subscript{max}=2500}=902$ and $2500-902=1598$, we iterate through $n_{i,\text{opt}}\superscript{U}\vert_{n\subscript{max}=1598}=674$ and $n_{i,\text{opt}}\superscript{U}\vert_{n\subscript{max}=924}=462$ (where the recursion stops as $924-462=462<n\subscript{min}\superscript{U}+n\subscript{min}\superscript{D}$). Hence, the optimal CLARQ schedule for $n\subscript{max}=2500$ is $\tilde{\mathbf{n}}\subscript{opt}=[902,674,462,462]$.

Then, we calculate the corresponding closed-loop error rate $\left(1-\mu_{(I),\text{d}}\superscript{loop}\right)$, as shown in Figs.~\ref{subfig:demo_closed_loop_err_sym} and \ref{subfig:demo_closed_loop_err_asym}. The error rate turns to be a monotonically decreasing and concave function of $n\subscript{max}$. Along with the increase of $I$ about $n\subscript{max}$, the error rate also appears segmented w.r.t. its first derivative, and remains quasi-log-linear to $n\subscript{max}$ in every individual segment. This behavior can be explained by the approximately log-quadratic feature of Q function: $Q(x)\approx \frac{1}{12}e^{ -x^2/2}+\frac{1}{4}e^{-2x^2/3}$ \cite{Chiani2003new}. It is trivial to prove that for sufficient values of $n$ that fulfill $\mathcal{C}^2n\gg d^2/n$, it approximately holds $\varepsilon(n)\sim e^{-n}$ for a single transmission attempt with blocklength $n$. The closed-loop error rate, according to \eqref{eq:closed_loop_reward}, is therefore quasi-log-linear to $n\subscript{max}$ for every fixed $I$.

{Furthermore, we are interested in the impact of dynamic retransmissions scheduled by CLARQ on the device energy consumption. In the context of power efficiency, it is common to focus on the uplink where power consumption is supported by batteries, instead of the downlink where the server is guaranteed with a prosperous power supply. Moreover, in the FBL regime, the signal processing generally consumes significantly less energy than the radio transmission does. Thus, with a consistent uplink transmission power level $P\superscript{U}=E\subscript{b}/\text{bit}$, the device energy consumption is determined by the blocklength usage in uplink. We investigate the UL energy consumption given different $n_\text{max}$ in both scenarios A and B, compute its expectation, upper bound (when $I$ UL attempts are made), and lower bound (when only one UL attempt is made), and compare them with the baseline of optimal one-shot scheme. As it can be observed from the results shown in Figs.~\ref{subfig:demo_power_efficiency_sym} and \ref{subfig:demo_power_efficiency_asym}, the energy consumption of optimal CLARQ equals that of the optimal one-shot scheme when $I=1$, since they are the same scheme in this case. For $I\geqslant 2$, CLARQ delivers an enhanced power efficiency over the baseline, in company with the improved closed-loop reliability.}

\subsection{Benchmarking Tests}
\begin{figure}[!htbp]
	\centering
	\includegraphics[width=.9\linewidth]{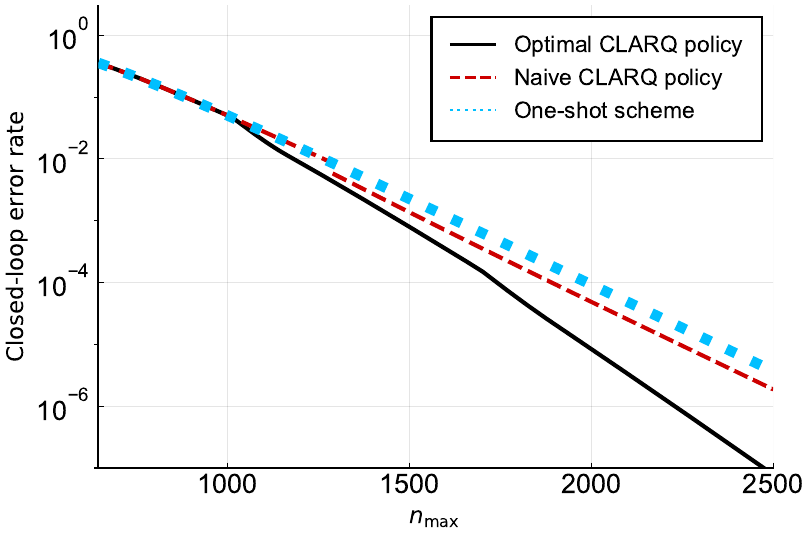}
	\caption{{Benchmark test w.r.t. closed-loop error rate as function of blocklength}}
	\label{fig:benchmarks_comp}
\end{figure}

To thoroughly evaluate the gains that can be achieved by optimal CLARQ, we compare its performance in scenario A to the following two baseline solutions:
\begin{enumerate}
	\item The one-shot FBL scheme, where no retransmission is scheduled, and the available blocklength is optimally allocated to the UL and DL slots so that $\varepsilon\superscript{U}=\varepsilon\superscript{D}$.
	\item A na\"ive CLARQ policy, where in every stage $i$ it takes the one-step-optimal allocation that forces $\varepsilon_i\superscript{U}=\varepsilon_i\superscript{D}$ (repeated optimal one-shot).
\end{enumerate}
As the results depicted in Fig.~\ref{fig:benchmarks_comp} show, when there is no resource to support any retransmission ($I=1$), all three methods have the same performance, since they are indeed suggesting the same scheme, i.e. equally allocating the blocklength to UL and DL slots. However, when the total blocklength $n\subscript{max}$ is sufficient to enable UL retransmission, the optimal CLARQ policy significantly outperforms the one-shot scheme, the performance gain concavely increases in a segmented quasi-linear fashion. In contrary, the gain provided by na\"ive CLARQ policy is negligible. Especially, at $n\subscript{max}=2500$ where the closed-loop air latency reaches the \SI{10}{\milli\second} constraint, the optimal CLARQ policy, with up to $I=3$ UL transmission opportunities, reduces the closed-loop error rate to the level of $7.8\times10^{-8}$, which shows a gain over $47$-fold in reference to the state-of-the-art (i.e. one-shot scheme) performance of $3.68\times10^{-6}$, while the na\"ive CLARQ policy is only capable to reduce the error rate by $50\%$. 

It is worth to remark that the applicability of our CLARQ performance model is limited to an upper bound of $n\subscript{max}$, as when the length of the longest transmission slot among all, i.e. $n_1\superscript{D}$, exceeds a certain level (which refers to 4000 according to~\cite{Makki2016wireless}), it shall be considered as the infinite blocklength case, and the error probability approximations in FBL regime \eqref{eq:os_ul_err_rate} and \eqref{eq:os_dl_err_rate} do not hold anymore.

\subsection{Sensitivity to Packet Size and SNR}
\begin{figure}[!htbp]
	\centering
	\includegraphics[width=.9\linewidth]{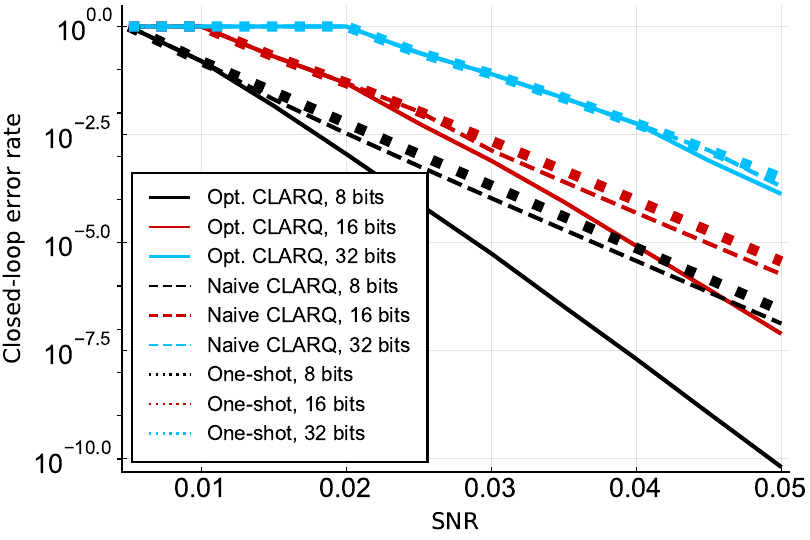}
	\caption{Sensitivities of optimal CLARQ and benchmarks to the SNR and packet size, where the channel is symmetric in UL/DL and $T=\SI{10}{\milli\second}$.}
	\label{fig:sensitivity_tests}
\end{figure}

Both the optimal CLARQ and the benchmarks are evaluated under different packet sizes and SNRs, where the channel is considered symmetric in UL/DL and $T$ is fixed at \SI{10}{\milli\second} (i.e. $n\subscript{max}=2500$), as shown in Fig.~\ref{fig:sensitivity_tests}. We can observe from the results that under all cases the optimal CLARQ generally holds a performance gain over the benchmarks, which is more significant with \begin{enumerate*}
	\item higher SNR, and
	\item smaller packet size.
\end{enumerate*}

\subsection{Sensitivity to Rayleigh Fading and Log-normal Shadowing}
In practical use scenarios, the wireless channels usually undergo fast Rayleigh fading and slow log-normal shadowing effects. To investigate the sensitivity of optimal CLARQ to the fading effects fading, we conducted Monte-Carlo tests, in every test the UL and DL channels are independently randomly generated, each with a base SNR level of \SI{10}{\dB}, affected by a 0-dB-mean log-normal random shadowing, and a Rayleigh fading.

\begin{figure}[!htbp]
	\centering
	\includegraphics[width=.9\linewidth]{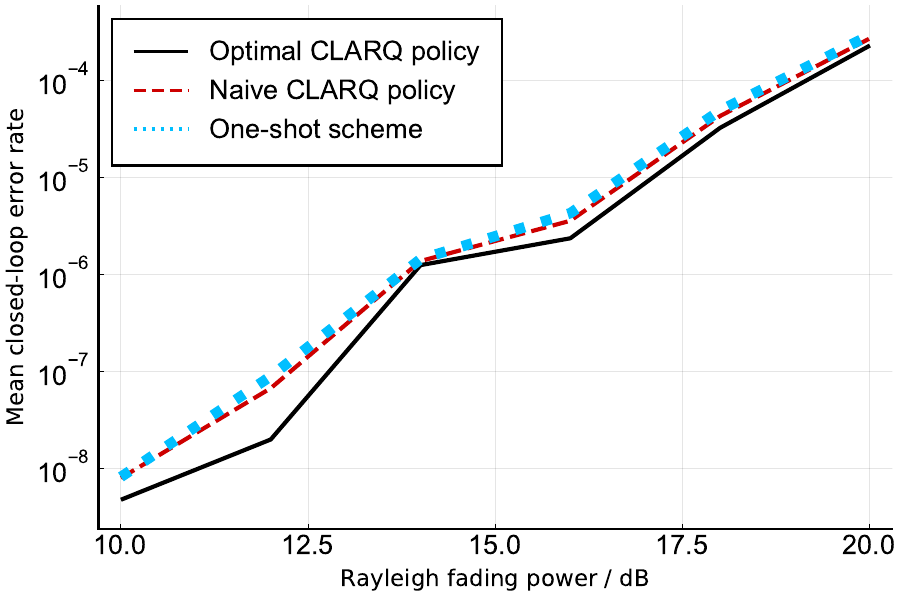}
	\caption{Sensitivity of optimal CLARQ to the Rayleigh fading.}
	\label{fig:sensitivity_rayleigh}
\end{figure}

First, we fixed the standard deviation of shadowing effect to \SI{3}{\dB}, and let the power of Rayleigh fading vary from \SI{10}{\dB} to \SI{20}{\dB}. For each specification we repeated the test 5000 individual times {so that in every test we numerically calculated} the closed-loop error probabilities of optimal CLARQ and both the baseline solutions. The results are depicted in Fig.~\ref{fig:sensitivity_rayleigh}, showing an increase of the closed-loop error rate along with the power of Rayleigh fading, and a consistent advance of the optimal CLARQ against the baselines.

\begin{figure}[!htbp]
	\centering
	\includegraphics[width=.9\linewidth]{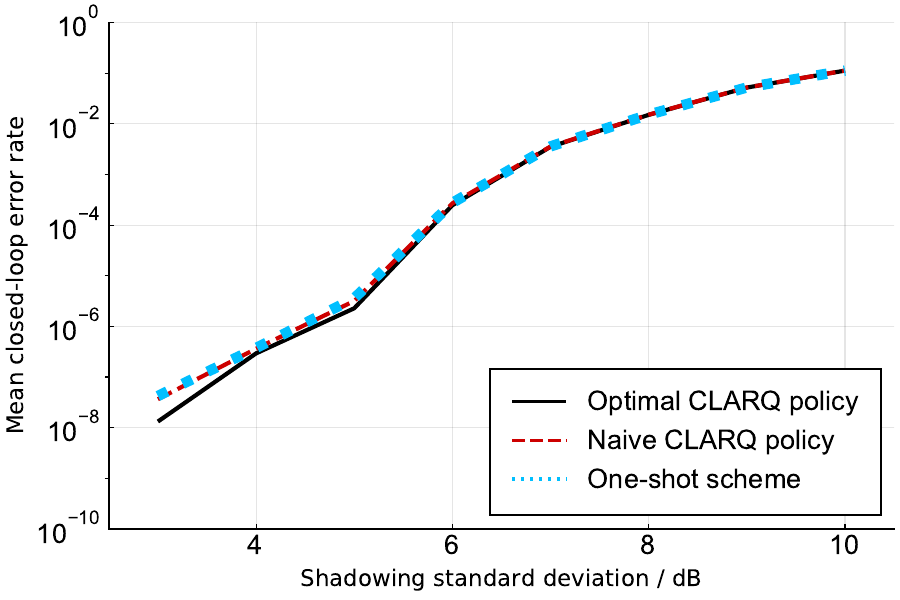}
	\caption{Sensitivity of optimal CLARQ to the shadowing.}
	\label{fig:sensitivity_shadowing}
\end{figure}

{Additionally, in our simulation settings we fixed the Rayleigh fading power to \SI{10}{\dB} and tested the optimal CLARQ policy assuming different standard deviations, ranging from \SI{3}{\dB} to \SI{10}{\dB}. The Monte-Carlo test was repeated $5000$ individual times under every specification and results are shown in Fig.~\ref{fig:sensitivity_shadowing}. In particular, we can notice a clear increase of the closed-loop error rate with the shadowing power, whereas the performance gain of the optimal CLARQ policy suffers when high shadowing power is considered. This is due to the high probability to have low SNR values.}

\subsection{Sensitivity to Look-Up-Table Resolution}
{As discussed in Section~\ref{subsec:computation_complexity}, practical deployment of CLARQ probably relies on the LUT-based implementation.} Obviously, the system performance will be determined by the SNR resolution of the LUT. To study the sensitivity of CLARQ to the LUT resolution, we carried out a exhaustive numerical simulation campaign. In every individual run, the UL and DL channels are independently randomly generated, each with a base SNR level of \SI{10}{\dB}, affected by a 0-dB-mean log-normal random shadowing with \SI{3}{\dB} standard deviation, and a Rayleigh fading with power of \SI{10}{\dB}. In each test, we numerically calculate the closed-loop error probabilities achieved by LUT-CLARQ with different LUT SNR intervals, as well as the optimal CLARQ performance. The average performances of different LUT implementations over 5000 runs of Monte-Carlo test are listed below in Table~\ref{tab:lut_test}, which shows that the ultra-reliable requirement can be fulfilled with a reasonably fine LUT resolution.
\begin{table*}[!htbp]
	\centering
	\caption{Benchmarking the LUT-CLARQ performance with different LUT resolutions}
	\label{tab:lut_test}
	\begin{tabular}{l|c|c|c|c|c|c}
		\toprule[2px]
		\textbf{SNR interval of LUT (dB)}&16&8&4&2&1&$\to 0$ (optimal CLARQ)\\\hline
		\textbf{Closed-loop error rate}&3.00e-3&4.98e-7&3.44e-8&2.98e-8&2.98e-8&2.31e-10\\
		\bottomrule[2px]
	\end{tabular}
\end{table*}

{
\section{Further Discussions}\label{sec:discussions}
\subsection{Impact of Feedback Loss}
For convenience of analysis, from Section~\ref{subsec:bellman_view} on we have taken the approximation $T_\text{f}\approx 0$, which is a common approach in the field of FBL information theory with ARQ mechanism. Generally, a non-zero feedback cost $T\subscript{f}>0$ leads to a loss in the overall blocklength utilization upon every retransmission, and may also reduce the maximal retransmission attempts $I$. Therefore, given a certain system specification, the gain of optimal CLARQ schedule over static ARQ/HARQ drops along with an increasing $T\subscript{f}$. Nevertheless, this does not violate the qualitative assertion of optimal CLARQ being superior over static ARQ/HARQ, since:\begin{enumerate}
	\item As we have proven in Lemma~\ref{lem:static_retr_opt}, where a generic $T\subscript{f}\geqslant 0$ is taken into account, the optimal static ARQ/HARQ strategy is always the optimal one-shot scheme.
	\item The optimal one-shot scheme is a special case of CLARQ scheme where $I=1$.
\end{enumerate}

\subsection{Interference Control in Multi-User System}
So far we have been discussing the performance of CLARQ in context of single-user systems, where the cross-user interference is not taken into account. For multi-user systems, it shall be remarked that the online duplex schedule may challenge the interference control. More specifically, the radio pattern of a device cannot be accurately predicted by its neighbors, so that it becomes impossible to establish a perfect interference-canceling link schedule. 

As a simple and conceptual demonstration, we investigate a minimal example where two devices are connected to the same server, both working in TDD mode, and the channel is symmetric in UL/DL for each device. We also consider the extreme case where the two devices are located distantly separated, being hidden nodes of each other, while the server is performing an imperfect beamforming. Thus, the devices are interfering each other when they both work in uplink, and when they both work in downlink, but not across UL/DL.

First we consider the system to work in OFDMA mode, with both users taking the static optimal one-shot scheme, i.e. equally split UL/DL time slots. In this case, an optimal cross-user schedule can be sketched as shown in Fig.~\ref{fig:cross_user_interference}, upper-left, where cross-user interference is mitigated. In contrast, when CLARQ is applied together with OFDMA, as Fig.~\ref{fig:cross_user_interference} illustrates in its bottom-left corner, since the lengths of uplink/downlink slots for both users are dynamically generated and therefore unknown in priori, an interference between two devices will become inevitable. However, when TDMA is applied instead of OFDMA, as shown in the upper-right part of Fig.~\ref{fig:cross_user_interference}, cross-user interference is again eliminated, while the overall available blocklength remains unchanged for both users. To this end, we evaluate CLARQ as more compatible with TDMA than with OFDMA regarding interference control.

\begin{figure}[!hbtp]
	\centering
	\includegraphics[width=.9\linewidth]{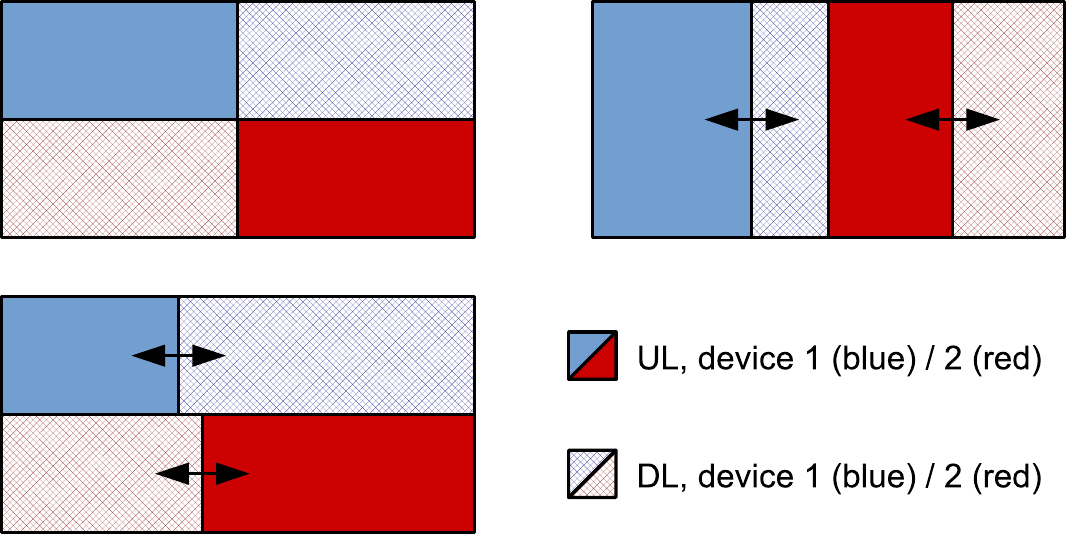}
	\caption{{Multi-user radio scheduling with optimal one-shot (upper-left), CLARQ+OFDMA (bottom-left), and CLARQ+TDMA (upper-right).}}
	\label{fig:cross_user_interference}
\end{figure}

\subsection{Adaptive Power Control}\label{subsec:apc}
In this work, we have been focusing on the optimal blocklength allocation under a consistent transmission power level. It is also a significant problem in the finite blocklength regime, however, to adaptively adjust the transmission power for a performance optimization in perspective of power control -- examples have been reported in~\cite{Makki2016wireless} and \cite{Makki2014green}.

A rational design of applying adaptive power control (APC) in the CLARQ protocol is to constraint not only the total blocklength $n\subscript{max}$, but also the maximal UL transmission energy $E\superscript{U}\subscript{max}$ available in every transmission frame $T$, and allowing the device to independently set the transmission power $P\superscript{U}_i$ for every individual UL attempt $i$. Thus, the device shall jointly optimize the blocklength usage and the energy consumption over its CLARQ scheme.

Unfortunately, adding the power variable and the energy consumption into the DP problem will dramatically increase the complexity. The reason is twofold: First, for finite blocklength, the joint convexity of power and blocklength%
, to the best of our knowledge, has never been proven, preventing us from simply decoupling the optimizations of power and blocklength from each other.
Second, the real-valued energy budget, unlike the discrete blocklength, can be arbitrarily divided. So the space of transmission power in every UL attempt $P\superscript{U}_i$ is a continuous range $[P\superscript{U}\subscript{min}, P\superscript{U}\subscript{max}]$ with infinite values, and its optimization is much more complex than that of the blocklength $n\superscript{U}\subscript{i}$ in a limited integer set.

Nevertheless, it is still possible to demonstrate the potential of APC in CLARQ with a simplified case study. We consider the reference scenario A and default system specifications as defined in Tab.~\ref{tab:sys_spec}, and the energy budget within a frame $T$ limited to the upper bound of CLARQ's UL energy consumption in its default mode, which is shown in Fig.~\ref{subfig:demo_power_efficiency_sym}.  Now we allow the device to set its transmission power level independently for each individual UL attempt: either in the default mode where $P\superscript{U}=E\subscript{b}/\text{bit}$, or in a high-power mode where $P\superscript{U}=1.25E\subscript{b}/\text{bit}$ (which therefore raises $\gamma\superscript{U}$ also by $25\%$). Due to the astronomical complexity of computation we only solved the power-blocklength joint optimum of this case with several $n\subscript{max}$ values, as listed in Tab.~\ref{tab:apc_case_study} with performance alongside the baseline of simple CLARQ without APC. It can be observed that APC does not only reduces the closed-loop error rate of CLARQ under the same constraint of energy budget, but also has the potential to improve the power efficiency. It becomes therefore worth to investigate the efficient solution of APC-CLARQ joint optimization in future.

\begin{table*}[!hbtp]
	\centering
	\caption{Adaptive power control enhances both reliability and power efficiency of CLARQ}
	\label{tab:apc_case_study}
	\begin{tabular}{l|c|c|c|c|c}
		\toprule[2px]
		\multirow{2}{*}{\textbf{Total blocklength}}& \textbf{Mean energy} & \textbf{Closed-loop} & \multirow{2}{*}{$(n\superscript{U}\subscript{1},n\superscript{D}\subscript{1},P\superscript{U}\subscript{1}/E\subscript{b})$} &\multirow{2}{*}{$(n\superscript{U}\subscript{2},n\superscript{D}\subscript{2},P\superscript{U}\subscript{2}/E\subscript{b})$} & \multirow{2}{*}{$(n\superscript{U}\subscript{3},n\superscript{D}\subscript{3},P\superscript{U}\subscript{3}/E\subscript{b})$} \\
		&\textbf{consumption}&\textbf{error rate}&&&\\\midrule[1px]
		\textbf{1200 (with APC)}&$2~266.6$&2.75e-3&$(434,766,1.25)$&$(349,417,1.25)$&N/A\\
		\textbf{1200 (without APC)}&$3~468.0$&8.65e-3&$(533,667,1)$&$(334,333,1)$&N/A\\\hline
		\textbf{1400 (with APC)}&$2~582.7$&4.40e-4&$(507,893,1.25)$&$(406,487,1.25)$&N/A\\
		\textbf{1400 (without APC)}&$4~008.0$&1.79e-3&$(604,796,1)$&$(398,398,1)$&N/A\\\hline
		\textbf{1600 (with APC)}&$2~830.6$&6.90e-5&$(562,1038,1.25)$&$(470,568,1.25)$&N/A\\
		\textbf{1600 (without APC)}&$4~552.0$&3.50e-4&$(675,925,1)$&$(463,462,1)$&N/A\\\hline
		\textbf{1800 (with APC)}&$2~719.4$&7.79e-6&$(523,1277,1.25)$&$(400,877,1.25)$&$(398,479,1.25)$\\
		\textbf{1800 (without APC)}&$5~912.0$&5.44e-5&$(692,1108,1)$&$(464,644,1)$&$(322,322,1)$\\
		\bottomrule[2px]
	\end{tabular}
\end{table*}

\subsection{HARQ Gain}
In this work, we have analyzed the performance of CLARQ in context of simple ARQ, i.e. if a transmitted packet fails to be decoded , it is simply discarded and never exploited in its future retransmissions. Since simple ARQ is known to delineate a tight upper bound for the error rate of all kinds of HARQ, it is reasonable to expect that advanced HARQ techniques, including type II and type III HARQ, shall be applied in CLARQ to incrementally reduce the error rate upon every extra retransmission, and therewith further enhance the performance.

From the performance, various HARQ implementations distinguish from each other mainly by the error rate in retransmissions. Upon the $i\superscript{th}$ retransmission, simple ARQ has
\begin{equation}\label{eq:err_rate_simple_arq}
	\varepsilon_{i,\text{ARQ}} = Q\left(\sqrt{\frac{n_i}{V}}\left(\mathcal{C}-\frac{d}{n_i}\right)\ln{2}\right),
\end{equation}
which applies for both $\varepsilon_{i}\superscript{U}$ and $\varepsilon_{i}\superscript{D}$ in Eq.~\eqref{eq:closed_loop_reward}. In comparison, for type II HARQ it is 
\begin{equation}\label{eq:err_rate_harq2}
	\varepsilon_{i,\text{HARQ- II}}= Q\left(\sqrt{\frac{n_{(i)}}{V}}\left(\mathcal{C}-\frac{d}{n_{(i)}}\right)\ln{2}\right),
\end{equation}
where $n_{(i)}=\sum\limits_{k=1}^in_k$. Generally, given the same total blocklength $n\subscript{max}$ and assuming negligible feedback loss, w.r.t. the overall error rate after all retransmissions, the one-shot scheme always outperforms all simple ARQ schemes with $I>1$, and is outperformed by type II HARQ, regardless of the specific assignment of $n_i$, i.e. $\varepsilon_{(I),\text{ARQ}}\geqslant\varepsilon_{\text{os}}\geqslant\varepsilon_{(I),\text{HARQ-II}}$. Nevertheless, since Lemma~\ref{lem:concavity}  holds for both \eqref{eq:err_rate_simple_arq} and \eqref{eq:err_rate_harq2}, the gain of CLARQ is valid for both simple ARQ or type II HARQ.

Yet it appears an open challenge, however, to prove the applicability of CLARQ with type III HARQ, where each retransmitted packet is self-decodable, so the precise expression of overall error probability is likely intractable and the convexity cannot be guaranteed.  This can be an interesting topic for the future work.

\section{Related Work}\label{sec:related}
Various approaches have been already proposed by literature to fulfill the reliability requirements of 5G URLLC-based use cases. One straightforward idea is to apply advanced resource allocation methods such that the radio resources can be more efficiently shared $i$) among devices of different classes~\cite{Abedin2019resource}, and $ii$) among different URLLC data packets~\cite{Anand2018resource}. In addition, it has been demonstrated that adaptive sub-carrier selection can also improve the link reliability in OFDM systems by raising the SNR and reducing adjacent-channel interference~\cite{Hamamreh2017ofdm}.
Alternatively, it is indicated by~\cite{Sachs20185g} that diversities in different domains (e.g., time, frequency, spatial) can play a key role in reliability enhancements. It is proved capable to significantly reduce packet errors by exploiting the multi-path spatial diversity in multi-hop networks, such as Cloud-RAN fronthauls~\cite{Mountaster2018reliable}, amplified-and-forward relaying networks~\cite{Tseng2019selective}, and aeronautical ad-hoc networks~\cite{Luo2018frudp}. 

Differing from the physical layer approaches~\cite{Abedin2019resource, Anand2018resource, Hamamreh2017ofdm}, our proposed data link layer method does not rely on cross-user resource allocation nor on specific multiplexing scheme. Indeed, CLARQ exploits the time diversity in an opportunistic fashion, and therefore is capable to apply in single-hop networks without making use of spatial-diversity-based methods~\cite{Mountaster2018reliable,Tseng2019selective,Luo2018frudp}.}

\section{Conclusion and Outlooks}\label{sec:conclusion}
In this paper, we have presented a novel TDD protocol, namely the CLARQ that allows to apply dynamic UL retransmission within a frame of limited length, so as to enable ARQ under strict latency constraints for higher closed-loop link reliability {and power efficiency}, in order to fulfill the extreme performance expectations of ultra-reliable use cases in wireless networks. We have analytically demonstrated some important features of CLARQ as a dynamic programming problem, and implemented an integer DP algorithm to efficiently solve its optimum. Our proposed methods have been verified by numerical results as significantly outperforming the state of the art, and capable of delivering ultra-high reliability together with low latency in practical Rayleigh channels.

{Following up this work, there are plenty potentials remaining for future study of ours and interested peers. In particular, novel cross-user interference control is needed as a key enabler for the application of CLARQ in OFDMA systems. It also remains an open and critical challenge, to efficiently solve the blocklength-power joint optimization to enable adaptive power control in CLARQ. Additionally, the applicability of CLARQ with type III HARQ is also an interesting research topic.}

\appendices
\section{Proof of Theorem~\ref{th:one_shot_opt}}\label{proof:one_shot_opt}
\begin{proof}
\revise{If there is no feasible combination of $n^U$ and $n^D$ that fulfills both constraints~\eqref{con:total_time} and \eqref{con:max_err_rate}, Problem (4) is considered infeasible, which can be usually represented by $(1-\varepsilon\superscript{U}\subscript{opt})(1-\varepsilon\superscript{D}\subscript{opt})=-\infty$ and $n\superscript{U}\subscript{opt},n\superscript{D}\subscript{opt}\in\emptyset$. Otherwise,} the scheduling can be optimized by finding the first deviate of \eqref{prob:tdd_prob_one_shot} under the constraint $n\superscript{U}+n\superscript{D}=\frac{T}{T\subscript{S}}=n_\Sigma$:
	\begin{equation}
	\begin{split}
	&\frac{\diff(1-\varepsilon\superscript{U})(1-\varepsilon\superscript{D})}{\diff n\superscript{U}}\\
	=&\left(\varepsilon\superscript{D}-1\right)\frac{\diff\varepsilon\superscript{U}}{\diff n\superscript{U}}+\left(\varepsilon\superscript{U}-1\right)\frac{\diff\varepsilon\superscript{D}}{\diff n\superscript{U}}=0.
	\end{split}
	\label{eq:extreme_one_shot}
	\end{equation}
	Since $\frac{\diff\varepsilon\superscript{U}}{\diff n\superscript{U}}<0$ while $\frac{\diff\varepsilon\superscript{D}}{\diff n\superscript{U}}=\frac{\diff\varepsilon\superscript{U}}{\diff \left(n-n\superscript{D}\right)}=-\frac{\diff\varepsilon\superscript{D}}{\diff n\superscript{D}}>0$, there exists a unique solution of  \eqref{eq:extreme_one_shot} that $\varepsilon\superscript{D}=\varepsilon\superscript{U}$. Recalling Eqs.~\eqref{eq:os_ul_err_rate} and \eqref{eq:os_dl_err_rate}, this requires
	\begin{equation}
		\begin{split}
			&\sqrt{\frac{n\superscript{U}\subscript{os}}{V\superscript{U}}}\left(\mathcal{C}\superscript{U}-\frac{d}{n\superscript{U}\subscript{os}}\right)\\
			=&\sqrt{\frac{n_\Sigma-n\superscript{U}\subscript{os}}{V\superscript{D}}}\left(\mathcal{C}\superscript{D}-\frac{d}{n_\Sigma-n\superscript{U}\subscript{os}}\right),
		\end{split}
	\end{equation}
	which can be reformatted into a cubic equation of $n\superscript{U}$ in standard form \eqref{eq:one_shot_opt}.
\end{proof}

\section{Proof of Lemma~\ref{lem:static_retr_opt}}\label{proof:static_retr_opt}
\begin{proof}
For all $i\in\{1,2\dots I\}$, \revise{according to}~\cite{Makki2014finite}:
\begin{align}
    \varepsilon\superscript{U}_{(i)}&\approx Q\left(\sqrt{\frac{n\superscript{U}_{(i)}}{V\superscript{U}}}\left(\mathcal{C}\superscript{U}-\frac{d}{n\superscript{U}_{(i)}}\right)\ln{2}\right),\\
    \varepsilon\superscript{D}_{(i)}&\approx Q\left(\sqrt{\frac{n\superscript{D}_{(i)}}{V\superscript{D}}}\left(\mathcal{C}\superscript{D}-\frac{d}{n\superscript{D}_{(i)}}\right)\ln{2}\right),
\end{align}
where $n\superscript{U}_{(i)}=\sum\limits_{j=1}^i n\superscript{U}_j$ and $n\superscript{D}_{(i)}=\sum\limits_{j=1}^i n\superscript{D}_j$ are the total blocklengths allocated to UL and DL, respectively. Especially, when $I=1$, it becomes the one-shot scheme. 

Now let $t\superscript{U}_{(i)}=n\superscript{U}_{(i)}T\subscript{S}+(i-1)T\subscript{f}$, it always holds that
\begin{align}
	 \varepsilon\superscript{U}_{(i)}\leqslant Q\left(\sqrt{\frac{t\superscript{U}_{(i)}}{V\superscript{U}T\subscript{S}}}\left(\mathcal{C}\superscript{U}-\frac{d}{t\superscript{U}_{(i)}}\right)\ln{2}\right),\label{eq:ul_err_rate_bound}
\end{align}
where the equality holds if and only if $(I-1)T\subscript{f}=0$. The same conclusion can be made for DL. Hence, given an arbitrary schedule $\left[n\superscript{U}_1, n\superscript{U}_2\dots n\superscript{U}_I, n\superscript{D}_1, n\superscript{D}_2\dots n\superscript{D}_I\right]$, we have the closed-loop reliability
\begin{equation}
	\mu\superscript{loop}_{(I), \text{s}}=\left(1-\varepsilon\superscript{U}_{(I)}\right)\left(1-\varepsilon\superscript{D}_{(I)}\right)\leqslant \mu\superscript{loop}\subscript{os}\left(t\superscript{U}_{(I)}\right),
\end{equation}
where 
\begin{equation}
	\begin{split}
		\mu\superscript{loop}\subscript{os}(t)
		=&\left[1-Q\left(\sqrt{\frac{t}{V\superscript{U}T\subscript{S}}}\left(\mathcal{C}\superscript{U}-\frac{dT\subscript{S}}{t}\right)\ln{2}\right)\right]\\
		\times&\left[1-Q\left(\sqrt{\frac{T-t}{V\superscript{D}T\subscript{S}}}\left(\mathcal{C}\superscript{D}-\frac{dT\subscript{S}}{T-t}\right)\ln{2}\right)\right]
	\end{split}
\end{equation}
is the closed-loop reliability of one-shot scheme with UL slot length $t$.
This implies that within $T$, the one-shot scheme outperforms all HARQ schedules in closed-loop reliability. Recalling Theorem \ref{th:one_shot_opt}, the optimum is achieved when
$n\superscript{U}_1=n\superscript{U}\subscript{os}, n\superscript{D}_1=n\superscript{D}\subscript{os}$.
\end{proof}

\section{Proof of Lemma~\ref{lem:concavity}}\label{proof:concavity}
\begin{proof}
	It has been shown in~\cite{Polyanskiy2010channel} that $\varepsilon\superscript{U}_i$ monotonically decreases w.r.t. $n_i\superscript{U}$. Additionally, when $\varepsilon\superscript{U}_i \leqslant 0.5$, it is also convex to $n_i$. Hence, the concavity can be shown by investigating the second derivative of $\mu_{i,\text{d}}\superscript{loop}$ as:
	\begin{equation}
	\begin{split}
		\frac{{{\diff^2}{\mu_{i,\text{d}}\superscript{loop}}}}{{\diff \left(n\superscript{U}_i\right)^2}}=&\left(\varepsilon _i\superscript{D}({T\superscript{D}_{i-1}} - {n_i\superscript{U}}T\subscript{S}) - 1\right)\frac{{{\diff^2}\varepsilon _i\superscript{U}({n_i\superscript{U}})}}{{\diff (n\superscript{U}_i)^2}}\\
		+&\left(\varepsilon _i\superscript{U}(n\superscript{U}_iT\subscript{S}) - 1\right)\frac{{{\diff^2}\varepsilon _i\superscript{D}({T\superscript{D}_{i-1}} - {n\superscript{U}_iT\subscript{S}})}}{{\diff (n\superscript{U}_i)^2}}\\
		+&2\frac{{\diff\varepsilon _i\superscript{U}({n\superscript{U}_i}T\subscript{S})}}{{\diff{n\superscript{U}_i}}}\frac{{\diff\varepsilon _i\superscript{D}({T\superscript{D}_{i-1}} - {n_i\superscript{U}}T\subscript{S})}}{{\diff{n_i\superscript{U}}}}	\leqslant 0
	\end{split}
	\end{equation}
\end{proof}
\vspace{-1cm}

\section{Proof of Theorem~\ref{th:uniq_solution}}\label{proof:uniq_solution}
\begin{proof}
	\revise{The CLARQ problem \eqref{eq:bellman_eq} seeks for an optimal dynamic resource allocation. This is the most classic type of dynamic programming problem, which has been deeply studied by the famous work of \emph{Bellman} \cite{Bellman_1952,Bellman_1954}, in deterministic and stochastic forms, respectively. Several significant results, such as the existence of unique optimum and some specific analytical characteristics of the optimum, have been provided by these literature. A full reproduction to their proofs specified to the CLARQ problem will be lengthy, so here we only transform \eqref{eq:bellman_eq} into the strict and generic form of the stochastic dynamic programming problem studied in \cite{Bellman_1954}, so that the features can be simply derived from some condition tests. We refer the readers interested in the detailed proofs to the original literature.}
	
	Let $p_1(n)=1-Q\left(\sqrt{\frac{n}{V}}(\mathcal{C}-r)\ln2\right)$, $p_2(n)=1-p_1(n)$, $g_1(n)=g_2(n)=h_2(n)=0$, $h_1(n)=p_1(n)$, $a_1=b_1=a_2=0$, $b_2=1$, we can construct the sequence
	\begin{align}
		f_1(n)&=\max\limits_{0\leqslant n_1\superscript{U}\leqslant n}\left\{p_1(n_1\superscript{U})\left[g_1(n_1\superscript{U})+h_1(n-n_1\superscript{U})\right]\right.\nonumber\\
		&\left.+p_2(n_1\superscript{U})\left[g_2(n_1\superscript{U})+h_2(n-n_1\superscript{U})\right]\right\},\\
		\underset{\forall i\in\{1,2\dots\}}{f_{i+1}(n)}&=\max\limits_{0\leqslant n_{i+1}\superscript{U}\leqslant n}\left\{p_1(n_{i+1}\superscript{U})\left[g_1(n_{i+1}\superscript{U})+h_1(n-n_{i+1}\superscript{U})\right.\right.\nonumber\\
		&\left.+f_i\left(a_1n_{i+1}\superscript{U}+b_1(n-n_{i+1}\superscript{U})\right)\right]\nonumber\\
		&\left.+p_2(n_{i+1}\superscript{U})\left[g_2(n_{i+1}\superscript{U})+h_2(n-n_{i+1}\superscript{U})\right.\right.\nonumber\\
		&\left.\left.+f_i\left(a_2n_{i+1}\superscript{U}+b_2(n-n_{i+1}\superscript{U})\right)\right]\right\}.
	\end{align}
	Thus, the \revise{CLARQ} problem \eqref{eq:bellman_eq} becomes seeking after $f_{I+1}\left(\frac{T}{T\subscript{S}}\right)$, \revise{which is a specified stochastic case of the problem studied in \cite{Bellman_1954}.} Easily we can derive that for all $j\in\{1,2\}$:
	\begin{itemize}
		\item Both $g_j$ and $h_j$ are continuous and wide-sense monotonically increasing in $\left[0,\frac{T}{T\subscript{S}}\right]$, and $g_j(0)=h_j(0)=0$;
		\item			$\sum\limits_{k=1}^2\sum\limits_{i=0}^{+\infty}p_k(a_k^i\frac{T}{T\subscript{S}})\left[g_j\left(a_k^i\frac{T}{T\subscript{S}}\right)+g_j\left(b_k^i\frac{T}{T\subscript{S}}\right)+h_j\left(a_k^i\frac{T}{T\subscript{S}}\right)\right.$\\$\left.+h_j\left(b_k^i\frac{T}{T\subscript{S}}\right)\right]<\infty$.
	\end{itemize}
	So the existence of a unique $\tilde{\mathbf{n}}_{I,\text{opt}}$ is ensured \revise{according to} \cite{Bellman_1954}.
	
	Furthermore, with sufficient $T$, the \revise{optima} $n_i\superscript{U}$ always \revise{fulfill} $n_i\superscript{U}>n\subscript{min}$ and therewith $\varepsilon_i\superscript{U}\leqslant 0.5$. Therefore:
	\begin{itemize}
		\item $\frac{\diff g_j}{\diff n\superscript{U}_i}\ge0, \frac{\diff^2 g_j}{{\diff n\superscript{U}_i}^2}\le0, \forall j\in\{1,2\}$ ($g_1=g_2=0$);
		\item $\frac{\diff h_1}{\diff n\superscript{U}_i}\ge0, \frac{\diff^2 h_1}{{\diff n\superscript{U}_i}^2}\le0$ (concavity from Lemma \ref{lem:concavity});
		\item $\frac{\diff h_2}{\diff n\superscript{U}_i}\ge0, \frac{\diff^2 h_2}{{\diff n\superscript{U}_i}^2}\le0$ ($h_2=0$);
		\item $b_1=a_1,b_2>a_2$,
	\end{itemize}
	which are addressed in \cite{Bellman_1954} as sufficient conditions for $n\superscript{U}_{1,\text{opt}}\ge n\superscript{U}_{2,\text{opt}}\ge\dots\ge n\superscript{U}_{I,\text{opt}}$ .
	For the last ($I\superscript{th}$) stage, we seek for the optimal $n_{I}\superscript{U}$ that maximizes $g(n_I\superscript{U})$ subjected to
		$n_I\superscript{U}+n_I\superscript{D}=n_{I-1,\text{opt}}\superscript{D}=\frac{T}{T\subscript{S}}-\sum\limits_{i=1}^{I-1}n_{i,\text{opt}}\superscript{U}$,
	which is achieved when $\varepsilon_I\superscript{U}\left(n_{I,\text{opt}}\superscript{U}\right)=\varepsilon_I\superscript{D}\left(n_{I,\text{opt}}\superscript{D}\right)$ as Theorem \ref{th:one_shot_opt} implies.
\end{proof}

\section{Proof of Corollary~\ref{cor:upper_lower_bounds_tdd}}\label{proof:upper_lower_bounds_tdd}
\begin{proof}
	The reward $\theta_i\left(n_i\superscript{U}\right)$ of the $i\superscript{th}$ UL transmission is maximized when its first derivative is zero. Recalling \eqref{eq:recursion_bellman} with $n^D_i=n^D_{i-1}-n^U_i$, it implies
	\vspace{-2mm}
	\begin{equation}
	(\varepsilon_i\superscript{D}-1)\frac{\diff \varepsilon_i\superscript{U}}{\diff n_i\superscript{U}}-(\varepsilon_i\superscript{U}-1)\frac{\diff\varepsilon_i\superscript{D}}{\diff n_i\superscript{D}}+\frac{\diff}{\diff n_i\superscript{U}}\left(\theta_{i+1}\varepsilon_i\superscript{U}\right)=0.\label{eq:theta_extreme}
	\end{equation}
	
	\vspace{-2mm}
	It is obvious that both $\theta_{i+1}$ and $\varepsilon_i\superscript{U}$ are non-negative and monotonically decreasing w.r.t. $n_i\superscript{U}$, therefore we have $\frac{\diff}{\diff n_i\superscript{U}}\left(\theta_{i+1}\varepsilon_i\superscript{U}\right)<0$. Moreover, since both $\varepsilon_i\superscript{U}$ and $\varepsilon_i\superscript{D}$ are bounded in $(0,1)$, and they are identically monotonically decreasing and convex in $n_i\superscript{U}$ and $n_i\superscript{D}$, respectively, it is trivial to see that \eqref{eq:theta_extreme} holds only if $\varepsilon_i\superscript{U}>\varepsilon_i\superscript{D}$. Therefore for the the optimal \revise{CLARQ} schedule we can assert
	\begin{equation}
		\varepsilon_{i,\text{opt}}\superscript{U}\ge\varepsilon_{i,\text{opt}}\superscript{D},\quad\forall i\in\{1,2\dots I\}.
	\end{equation}
	Especially, for TDD systems where UL and DL channels are symmetric, i.e. $\mathcal{C}\superscript{U}=\mathcal{C}\superscript{D}$, $V\superscript{U}=V\superscript{D}$, $n\subscript{min}\superscript{U}=n\subscript{min}\superscript{D}=n\subscript{min}$, this implies that
	\begin{equation}
		n_{i,\text{opt}}\superscript{U}\leqslant n_{i,\text{opt}}\superscript{D},\quad\forall i\in\{1,2\dots I\},\label{eq:ul_le_dl}
	\end{equation}
	where the equity only holds for the last stage ($i=I$).

	Consider an arbitrary $I$-stage schedule with $n_{I}\superscript{D}\ge2n\subscript{min}$ and an overall closed-loop reliability $\mu_0$. Obviously, by adding an extra $(I+1)\superscript{th}$ stage to the schedule with $n_{I+1}\superscript{U}=n_{I+1}\superscript{D}=\frac{1}{2}n_{I}\superscript{D}\ge n\subscript{min}$, it always leads to a better reward
	$\mu_{(I+1),\text{d}}\superscript{loop}=\mu_0+\left(1-\varepsilon_{I+1}\superscript{U}\right)\left(1-\varepsilon_{I+1}\superscript{D}\right)\prod\limits_{i=1}^I\varepsilon_i\superscript{U}>\mu_0$,
	so that the original $I$-stage schedule cannot be the optimum. Therefore, the optimal schedule must fulfill
	$n_{I,\text{opt}}\superscript{D}<2n\subscript{min}$.
	Additionally, taking into account the minimal blocklength $n\subscript{min}$ for all UL/DL transmissions, recalling \eqref{eq:blocklength_iteration} and \eqref{eq:ul_le_dl}:
	\vspace{-3mm}
	\begin{align}
	n\subscript{min}\le& n_{I,\text{opt}}\superscript{U}=n_{I,\text{opt}}\superscript{D}<2n\subscript{min}\nonumber\\
	2n\subscript{min}\le&n_{I-1,\text{opt}}\superscript{D}<4n\subscript{min}\nonumber\\
	n\subscript{min}\le&n_{I-1,\text{opt}}\superscript{U}<4n\subscript{min}\nonumber\\
	3n\subscript{min}\le&n_{I-2,\text{opt}}\superscript{D}<8n\subscript{min}\nonumber\\
	n\subscript{min}\le&n_{I-2,\text{opt}}\superscript{U}<8n\subscript{min}\nonumber\\
	\dots\nonumber&\\
	n\subscript{min}\le&n\superscript{U}_{i,\text{opt}}<2^{I-i+1}n\subscript{min}\nonumber\\
	(I-i+1)n\subscript{min}\le&n\superscript{D}_{i,\text{opt}}<2^{I-i+1}n\subscript{min}\nonumber
	\end{align}
\end{proof}
\vspace{-6mm}

\section{Proof of Corollary~\ref{cor:number_stages}}\label{proof:number_stages}
\begin{proof}
	$\frac{T}{T\subscript{S}}=n_{1,\text{opt}}\superscript{U}=n_{1,\text{opt}}\superscript{D}$, {so that} Corollary~\ref{cor:upper_lower_bounds_tdd} implies 
	$(I+1)n\subscript{min}\le\frac{T}{T\subscript{S}}< 2^{I+1}n\subscript{min}$.
	Hence,
	\begin{align}
	&(I+1)T\subscript{min}\leqslant T < 2^{I+1}T\subscript{min}\\
	&\log_2\left(\frac{T}{T\subscript{min}}\right)-1< I\le\frac{T}{T\subscript{min}}-1
	\end{align}
\end{proof}
\vspace{-5mm}

\ifCLASSOPTIONcaptionsoff
  \newpage
\fi


%
\vfill

\begin{IEEEbiography}[{\includegraphics[width=1in,height=1.25in,clip,keepaspectratio]{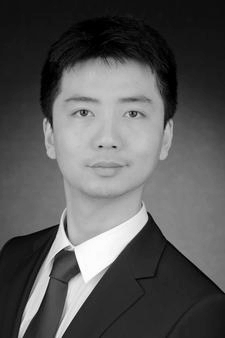}}]{Bin Han} (M'15--SM'21)
	received in 2009 his B.E. degree from Shanghai Jiao Tong University, M.Sc. in 2012 from Technical University of Darmstadt, and in 2016 the Ph.D. degree from Karlsruhe Institute of Technology. Since July 2016 he has been with University of Kaiserslautern as Postdoctoral Researcher and Senior Lecturer, researching in the broad area of wireless communication and networking. He is the author of over 40 research papers and book chapters, and have participated in multiple EU collaborative research projects. He serves as an Editor of \emph{Network (MDPI)} and \emph{Electronics (MDPI)}, and as a TPC member of GLOBECOM, EuCNC, and European Wireless.
\end{IEEEbiography}


\begin{IEEEbiography}[{\includegraphics[width=1in,height=1.25in,clip,keepaspectratio]{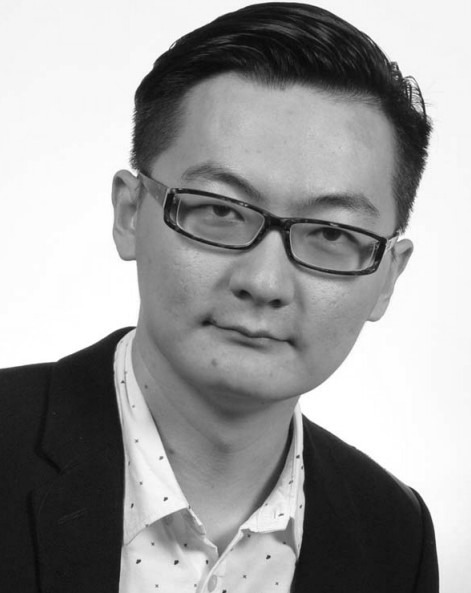}}]{Yao Zhu} (S'19) received the B.S. degree in electrical engineering from the University of Bremen, Bremen, Germany, in 2015, and the master’s degree in information technology and computer engineering from RWTH Aachen University, Aachen, Germany, in 2018. He is currently pursuing the Ph.D. degree with the ISEK Research Group, RWTH Aachen University. His research interests include ultra-reliable and low-latency communications, and mobile edge networks.
\end{IEEEbiography}


\begin{IEEEbiography}[{\includegraphics[width=1in,height=1.25in,clip,keepaspectratio]{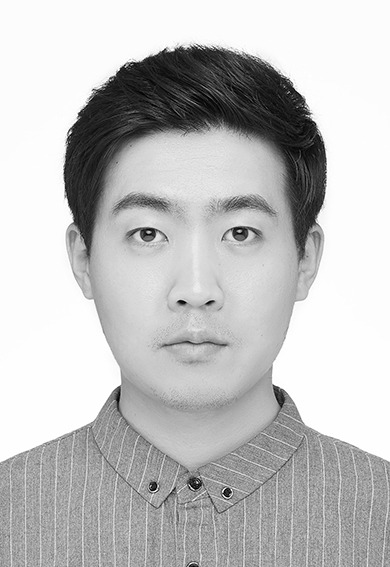}}]{Muxia Sun} received in 2010 his B.Sc. degree from South China University of Technology (SCUT), M.Sc. in 2012 \& 2013 from Universit\'e de Nantes and SCUT, respectively, and the Ph.D. degree in 2019 from Universit\'e Paris-Saclay. Since 2020 he has been with Tsinghua University as Postdoctoral Researcher in the Department of Industrial Engineering. His current research interests include reliability assessment and optimization of industrial \& communication systems, robust optimization, and approximation algorithm design.
\end{IEEEbiography}


\begin{IEEEbiography}[{\includegraphics[width=1in,height=1.25in,clip,keepaspectratio]{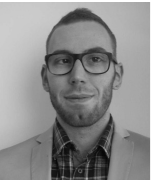}}]{Vincenzo Sciancalepore} (S'11--M'15--SM'19) received his M.Sc. degree in Telecommunications Engineering and Telematics Engineering in 2011 and 2012, respectively, whereas in 2015, he received a double Ph.D. degree. Currently, he is a senior 5G researcher at NEC Laboratories Europe GmbH in Heidelberg, focusing his activity on network virtualization and network slicing challenges. He is currently involved in the IEEE Emerging Technologies Committee leading the initiatives on SDN and NFV. He was also the recipient of the national award for the best Ph.D. thesis in the area of communication technologies (Wireless and Networking) issued by GTTI in 2015. He is an Editor of \emph{IEEE Transactions on Wireless Communications}.
\end{IEEEbiography}


\begin{IEEEbiography}[{\includegraphics[width=1in,height=1.25in,clip,keepaspectratio]{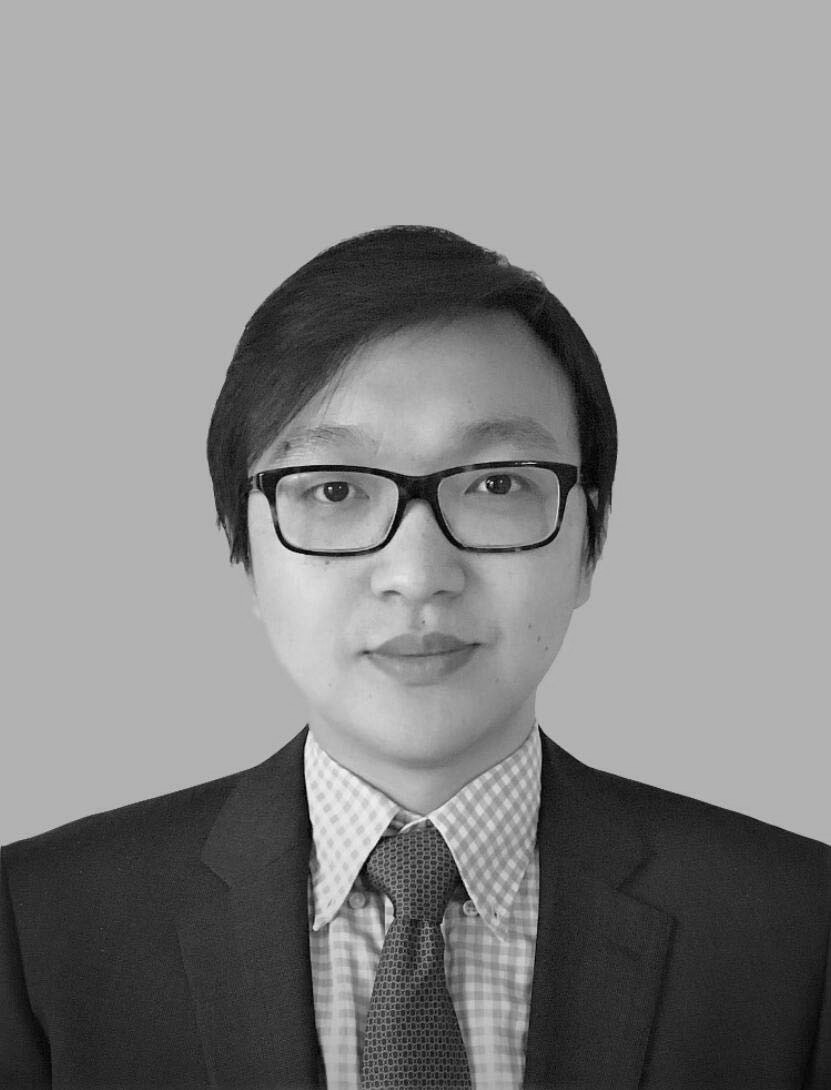}}]{Yulin Hu}(S11-M15-SM18) received his M.Sc.E.E. degree from USTC, China, in 2011. In Dec. 2015 he received his Ph.D.E.E. degree (Hons.) from RWTH Aachen University where he was a postdoctoral Research Fellow since Jan. to Dec. in 2016. He was a senior researcher and team leader in ISEK research Area at RWTH Aachen University. From May to July in 2017, he was a visiting scholar in Syracuse University, USA. He is currently a professor with School of Electronic Information, Wuhan University, and an adjunct professor with ISEK research Area, RWTH Aachen University. His research interests are in information theory, optimal design of wireless communication systems. He has been invited to contribute submissions to multiple conferences.  He was a recipient of the IFIP/IEEE Wireless Days Student Travel Awards in 2012. He received the Best Paper Awards at IEEE ISWCS 2017 and IEEE PIMRC 2017, respectively. He served as a TPC member for many conferences. He was the lead editor of the Urllc-LoPIoT spacial issue in \emph{Physical Communication}, and the organizer and chair of two special sessions in IEEE ISWCS 2018 and ISWCS 2020.  He is currently serving as an editor for \emph{Physical Communication (Elsevier)}, \emph{EURASIP Journal on Wireless Communications and Networking}, and \emph{Frontiers in Communications and Networks}. 
\end{IEEEbiography}


\begin{IEEEbiography}[{\includegraphics[width=1in,height=1.25in,clip,keepaspectratio]{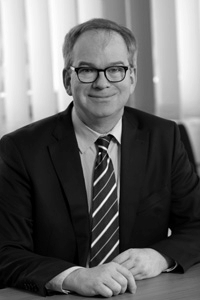}}]{Hans D. Schotten} (S'93--M'97)received the Ph.D. degree from the RWTH Aachen University of Technology, Germany, in 1997. From 1999 to 2003, he worked with Ericsson. From 2003 to 2007, he worked with Qualcomm. He became the Manager of a R\&D Group, a Research Coordinator for Qualcomm Europe, and the Director for Technical Standards. In 2007, he accepted the offer to become the Full Professor with the University of Kaiserslautern. In 2012, he became a Scientific Director of the German Research Center for Artificial Intelligence (DFKI) and the Head of the Department for Intelligent Networks. He served as the Dean of the Department of Electrical Engineering, University of Kaiserslautern from 2013 until 2017. He has authored more than 200 papers and participated over 40 European and national collaborative research projects. Since 2018, he has been the Chairman of the German Society for Information Technology and a Member of the Supervisory Board of the VDE.
\end{IEEEbiography}


\end{document}